\documentclass[conference]{IEEEtran}
\IEEEoverridecommandlockouts
\usepackage{cite}
\usepackage{amsmath,amssymb,amsfonts}
\usepackage{algorithmic}
\usepackage{graphicx}
\usepackage{textcomp}
\usepackage{xcolor}
\usepackage{hyperref}
\def\BibTeX{{\rm
B\kern-.05em{\sc i\kern-.025em b}\kern-.08em
    T\kern-.1667em\lower.7ex\hbox{E}\kern-.125emX}}
\begin{document}

%\title{A DGCNN-Based Approach to the Science4Cast Competition*\\

%\thanks{Identify applicable funding agency here. If none, delete this.}
%}

\title{A Method to Predict Semantic Relations on Artificial Intelligence Papers\\

%\thanks{Identify applicable funding agency here. If none, delete this.}
}

\author{\IEEEauthorblockN{1\textsuperscript{st} Francisco Andrades}
\IEEEauthorblockA{\textit{Department of Informatics} \\
\textit{Federico Santa Mar\'ia Technical University}\\
Santiago, Chile \\
francisco.andrades@sansano.usm.cl}

\and
\IEEEauthorblockN{2\textsuperscript{nd} Ricardo \~Nanculef}
\IEEEauthorblockA{\textit{Department of Informatics} \\
\textit{Federico Santa Mar\'ia Technical University}\\
Santiago, Chile \\
jnancu@inf.utfsm.cl}
}

\maketitle

\begin{abstract}
Predicting the emergence of links in large evolving networks is a difficult task with many practical applications. Recently, the Science4cast competition has illustrated this challenge presenting a network of 64.000 AI concepts and asking the participants to predict which topics are going to be researched together in the future. In this paper, we present a solution to this problem based on a new family of deep learning approaches, namely Graph Neural Networks. 

The results of the challenge show that our solution is competitive even if we had to impose severe restrictions to obtain a computationally efficient and parsimonious model: ignoring the intrinsic dynamics of the graph and using only a small subset of the nodes surrounding a target link. Preliminary experiments presented in this paper suggest the model is learning two related, but different patterns: the absorption of a node by a sub-graph and union of more dense sub-graphs. The model seems to excel at recognizing the first type of pattern. 

%This document is a model and instructions for \LaTeX.
%This and the IEEEtran.cls file define the components of your paper [title, text, heads, etc.]. *CRITICAL: Do Not Use Symbols, Special Characters, Footnotes, 
%or Math in Paper Title or Abstract.\\a\\a\\a\\a\\a
\end{abstract}


\section{Introduction}
%hablar más en general del problema de predecir links 
%terminar diciendo que el desafío XXX busca incentivar/expandir la investigación en esa área
%descubrir muy brevemente el enfoque y los resultados
In recent years, there has been an increasing interest on learning from graph structured data \cite{2018}. As graphs can be used to represent data in a diverse range of disciplines, problems like graph or node classification have a wide range of applications.

One of the main topics in graph learning is the problem of \emph{Link Prediction}: given a graph at some time $t$, the goal is to predict the formation of new links in future states of the graph. Many approaches have been developed throughout the years. Graph Kernels \cite{JMLR:v12:shervashidze11a}, Node embeddings \cite{DBLP:journals/corr/GroverL16} and traditional Machine Learning methods \cite{Krenn1910} have all been widely used with good performance. More recently, Deep Learning has allowed to achieve new, improved results \cite{DBLP:journals/corr/abs-1812-04202}. 

The Science4Cast challenge seeks to encourage research on Link Prediction \cite{Krenn1910}. It proposes a new benchmark dataset that contains semantic information on scientific papers and can be used to forecast emerging research topics on the Artificial Intelligence field.

The solution presented in this paper is based on a new, specialized family of Neural Networks, namely Graph Neural Networks (GNN) \cite{2021}. Experimental results show that the method can achieve competitive performance using only a subset of the nodes surrounding a target link and ignoring valuable information about the dynamics of the network (which is used by other methods). These simplifications lead to a solution with \emph{low memory usage} that can be used without specialized hardware. 

Preliminary experiments we present in this paper suggest the model is learning two related, but different patterns. The first is a kind of \textit{cold-start} scenario in which a node is absorbed by a larger sub-graph. The second is the union of two dense sub-graphs. The model seems to excel at recognizing the asymmetric cold-start case.
%More importantly, we present evidence that suggests a potential divergence on the problem to be solved, according to a proposed categorization of each link. We speculate that there are two problems that are fundamentally different and the network learns, in fact,  different functions to solve them.

The rest of this paper is structured as follows. In the next section we provide an overview of the Science4cast challenge. In Section \ref{sec:background}, we introduce the problem of link prediction providing a brief summary of the background used to obtain our method and the technique used as baseline. In Section \ref{sec:method} we present our solution, stressing the restrictions we had to impose to obtain an efficient method. Section \ref{sec:results} presents numerical results on the test data made available to the participants of the challenge. Section \ref{sec:conclusions} concludes the work with some final remarks.   
\section{The Challenge}\label{sec:challenge}
%describir el desafío en detalle
The massive amount of scientific papers (and its accelerated growth) allows for a detailed analysis on the behavior and evolution of scientific disciplines with an unprecedented level of sophistication \cite{Krenn1910} \cite{doi:10.1126/science.aao0185}.

This is the motivation behind the Science4Cast Challenge, that seeks to predict emerging trends in AI.

To achieve this, a semantic network obtained from an
intelligent analysis on a large corpus of AI papers is presented to the community \cite{Krenn1910}. In the resulting graph, nodes represents semantic concepts relevant to the field and edges represents the co-occurrence of two concepts in the same paper.

The challenge is as follows. Given the graph formed until the year 2017, predict the probability of link formation for different pairs of nodes by the year 2020. This is equivalent to predict which pairs of concepts are going to be researched together during the upcoming 3 years.

From a technical perspective, the challenge can be viewed as a Link Prediction problem on a partially dynamic graph: new edges can be added between existing vertices on a temporal dimension.

A number of useful applications can be derived from this challenge: from an assistant system which helps researchers to choose their subjects, to the meta-study of science dynamics and evolution; passing through novel approaches to the link prediction problem.

\subsection{Dataset}

The network given for the challenge contains information about the co-occurrence of concepts in the AI field since 1994 until 2017. It has around 64000 nodes, each representing a concept addressed in a paper of AI. A list of edges and their creation date (discretized in days, since 01/01/1990) is used to represent the network. There are 7652945 edges formed by the end of 2017. A list of $1E6$ vertex pairs without links by 2017 is designated to make the predictions for 2020. These potential links don't have any restrictions and were chosen completely at random. 

An older snapshot of the network is given for testing/validation. A list of $1E6$ vertex pairs without links by the end of 2014 is designated as a standardized test set to make the predictions for 2017. The ground-truth for these test pairs is explicitly provided. Most of the results reported in this paper were obtained on this standardized test set.\

\begin{table}[h!]
\centering
\begin{tabular}{|c c c c c|} 
 \cline{4-5}
\multicolumn{3}{c}{} & \multicolumn{2}{|c|}{\# of degree zero nodes} \\
\hline
   & \# Nodes & \# Edges & 1 & 2 \\ [0.5ex] 
 \hline\hline
 2014 & 64719 & 2278611 & 49.4\%  & 30.9\% \\ 
 \hline
 2017 & 64719 & 7652945 & 31.8\%  & 3.9\% \\
 \hline 
\end{tabular}
\medskip
\caption{Descriptive Statistics of the Standardized Test Set (2014) and Challenge Set (2017): Number of nodes, Number of edges, Percentage of test pairs in which exactly 1 node has degree 0, Percentage of test pairs in which both nodes have degree 0}
\label{table:1}
\end{table}

%Table \ref{table:1}

The semantic meaning of each node is hidden for the participants. In fact, it is missing any type of feature information about nodes and edges. Therefore, only the topology can be used to address the challenge. A notorious problem with this is that there is a significant amount of test pairs where at least one of the nodes has degree 0, which means that there is no information about their relationships with other nodes. This is similar to the cold-start problem in collaborative filtering, a problem that could be addressed using additional metadata. 

For this paper, potential links will be categorized under 3 types: 1) Type 0: Links where both nodes has a nonzero degree, 2) Type 1: Links where exactly 1 node has a nonzero degree and 3) Type 2: Links where both nodes have a degree of 0.

It is worth noting that some edges can be repeated if the same pair of concepts is found in different papers, which results in a weighted undirected graph. 

\section{Background}\label{sec:background}

A variety of methods have been investigated to learn from graph data. Graph kernels, Node embeddings and traditional Machine Learning methods have been widely used. Recently, Deep Learning has allowed for breakthroughs on different areas of graph learning.

\subsection{Machine Learning based on Feature Engineering}

In this approach, a feature vector is first designed using e.g. metrics from graph theory such as degree, CN, and Katz index. Then, a traditional supervised algorithm is trained. The method in \cite{Krenn1910} is proposed in the Science4cast Challenge as a competitive baseline referred to as \emph{MK's Baseline}. The feature vector was designed using $15$ metrics that captures information about a pair of nodes along the temporal dimension: Degree of both nodes in the current year and previous two years, total number of shared neighbours of each node in the current year and previous two years, number of shared neighbours in current year an previous two years. A 2-layer Fully Connected Neural Network is then trained on the representation.

A modified version of the MK's Baseline is used in this paper to compare the results on the standardized test set. All the Type 2 vertex pairs were excluded from the input data and were assigned a prediction 0.

\subsection{Graph Neural Networks}

Graph Neural Networks (GNN) are a specialized family of Neural Networks that can receive the graph directly as an input \cite{kipf2017semisupervised} \cite{li2017gated} \cite{2018graph} \cite{you2021design}. Based on message passing mechanisms, the main idea is to learn a latent representation for each node aggregating the information about the neighbours using learnable parameters. As GNN grow in popularity, many layers have been proposed to implement this idea. A GCN is characterized by having a layer that resembles a convolution operation.

\paragraph{GCN Layer}: This is the standard layer for graph convolution operation \cite{kipf2017semisupervised}. For a given undirected graph $G$ with $N$ nodes, adjacency matrix $A \in R^{N \times N}$, and node features matrix $X \in R^{N \times c}$, the layer computes the following propagation
$$H\textsuperscript{(l+1)} = \sigma{(\tilde{D}^{-\frac{1}{2}}\tilde{A}\tilde{D}^{-\frac{1}{2}}H\textsuperscript{(l)}W\textsuperscript{(l)} )} \, , $$\

where $H\textsuperscript{(l)} \in R^{N \times D}$ is the matrix of activations of layer $l$ ($H^0 = X$), $\tilde{A} =  A \ + \ I_N$ is the adjacency matrix of the graph $G$ with added self connections, $I_N$ is the $N \times N$ identity matrix, $\tilde{D}_{ii} = \sum_{j} \tilde{A}_{ij}$, $W\textsuperscript{(l)}$ is a layer-specific trainable weight matrix, and $\sigma{(\cdot)}$ is a non linear activation function (usually $\mbox{tanh}(\cdot)$).

Many specialized architectures based on graph convolutional layers for specific problems have been proposed. In \cite{Zhang_Cui_Neumann_Chen_2018}, an architecture for graph classification was presented, namely DGCNN. In a DGCNN, a stack of graph convolutional layers is used to extract  latent representations of the nodes. A novel SortPooling layer takes as an input the concatenation of every GCN layer output, sorts the nodes representations in a consistent order and standardizes the output to a fixed size $k \times \sum c_i$, where $c_i$ is the number of channels of the graph convolutional layer $i$. The parameter $k$ filter the number of nodes considered in the layer's output, therefore, should be fine tuned accordingly.

A stack of 1D convolutional layers consumes the flattened output of the SortPooling layer, and a few fully-connected layers completes the model.

%First, we develop a novel �-decaying heuristic theory. The theory unifies a wide range of heuristics in a single framework, and proves that all these heuristics can be well approximated from local subgraphs.
\medskip
\paragraph{Link prediction using GNN}

For link prediction, \cite{zhang2018link} proposed the SEAL framework. Let's define:
\begin{itemize}
\item The \textit{h-hop \ enclosing \ sub-graph}. Given a graph $G = (V,E)$ and two nodes $(x,y) \in V$, the h-hop enclosing sub-graph for $(x,y)$ is the subgraph $G^h_{x,y}$ induced from $G$ by the set of nodes $\{ i \ | \ d(i, x) \le h \ or \ d(i, y) \le h \}$, where $d(a, b)$ denotes the length of the path from $a \in V$ to $b \in V$.

\item A \textit{h-order heuristic}. A heuristic score of node similarity which requires knowledge up to h-hop neighborhood of the target nodes.
\end{itemize}

\medskip
Using these definitions, its easy to prove the following theorem.

\textit{Theorem 1. Any h-order heuristic for $(x,y)$ can be accurately calculated from $G^h_{x,y}$.}

Moreover, they propose a novel \textit{$\gamma$-decaying  heuristic theory} that unifies a wide range of heuristics in a single framework. It can be proved that any high-order $\gamma$-decaying heuristic can be approximated by the \textit{h-hop enclosing sub-graph} with an error that, under certain conditions, decreases at least exponentially with $h$. Lastly, they demonstrate that most well-known heuristics for link prediction can be represented as \textit{$\gamma$-decaying heuristics}.

According to these results, the SEAL framework addresses a link prediction problem as the task of classifying the sub-graphs enclosing the target nodes. This requires 3 stages:
\begin{itemize}
    \item Enclosing sub-graphs extraction.
    \item Node information matrix construction $X$.
    \item GNN training.
\end{itemize}

\textit{Node Information Matrix Construction.} This step is often crucial for training an accurate link prediction model. The node information matrix $X$ in SEAL usually accommodate three types of features: explicit node features, node embeddings, and structural labels. Structural labels encode the role of each node in the topology. The center nodes $x$ and $y$ are the target nodes between which the link is located. Nodes with different relative positions to the center have different structural importance to the link. 

\cite{zhang2018link} proposed the \textit{Double-Radius Node Labeling (DRNL)} method $f_l:V \rightarrow N$, which assigns an integer label $f_l(i)$ to every node $i$ in the enclosing sub-graph. $f_l(i)$ takes the form
$$f_l(i) = 1 + min(d_x , d_y ) + (\frac{d}{2})[(\frac{d}{2}) + (d \mathbin{\%} 2) - 1]\, ,$$
where $d_x := d(i, x)$, $d_y := d(i, y)$, $d := d_x + d_y$ , $(\frac{d}{2})$ and $(d \mathbin{\%} 2)$ are the integer quotient and
remainder of $d$ divided by 2, respectively. 

\section{Proposed Method}\label{sec:method}

There is a variety of decisions to be made on how the challenge should be posed. The method presented in this paper has 2 main stages: 1) Data usage strategy, and 2) SEAL - DGCNN training.

%\begin{itemize}
%    \item Data Usage Strategy.
%    \item SEAL - DGCNN Training.
%\end{itemize}

\subsection{Data Usage Strategy}

As the challenge's goal is to use a graph formed until a determined year to predict the existence of a potential link 3 years in the future, it makes sense to prepare training data with exactly the same condition. For a given dataset, let's define:
\begin{itemize}
    \item \textit{Input year: The last year until the graph is formed.}
    \item \textit{Target year: The year for which is required to make predictions.}
\end{itemize}

In the challenge, $Target \ year  = Input \ year + 3$. For a given pair $(Input \ year, Target \ year)$ the training data will be created using: 
\begin{itemize}
    \item $Input \ year \ training = Input \ year - 3$.
    \item $Target \ year \ training = Input \ year$.
\end{itemize}

Therefore, the training inputs for the model will include node pairs $(a,b)$ for which no links exist in the graph by the end of $input \ year \ training$. The training labels/targets will be $+1$ for node pairs for which a link  has been observed by $target \ year \ training$ and $0$ for node pairs for which no link has been observed by $target \ year \ training$.

The training pairs $(a,b)$ were chosen almost at random. There were only 2 restrictions made:
\begin{itemize}
    \item The frequencies of the two classes were forced to be roughly the same, that is, an stratified sampling was applied. 
    \item There was a restriction for considering a training pair as eligible: as there is absolutely no information about nodes with degree 0, only type 1 or type 2 pairs $(a,b)$ were selected.
\end{itemize}

It is worth noting that the proposed method uses only the final state of the graph as an static input for the network, without considering any form of information on the temporal dimension. 

\subsection{SEAL - DGCNN Training}

The SEAL Framework based on a DGCNN network is proposed to predict link formation, i.e., for each input pair, the \textit{h-hop enclosing sub-graph} is extracted and that graph is classified by a DGCNN. 
For computational constraints, only \textit{1-hop enclosing sub-graphs} were used and $65\%$ of it's nodes were dropped at random.

For each sub-graph, the feature matrix $X$ is constructed as a one-hot encoding of the DRNL node labeling method described in the previous section. 

The DGCNN architecture used for the challenge consists of $4$ stacked GCN Layers with $128$ channels/filters each, a $1$ channel GCN Layer, a SortPooling Layer, $2$ 1D Convolutional Layers with $64$ and $128$ channels respectively, a MaxPooling Layer,  and $2$ 1D Convolutional Layers with $128$ and $64$ channels respectively, followed by $3$ fully-connected layers with $256$, $128$ and $1$ neuron each.

The value of $k$ for the Sort Pooling Layer changes according to the dataset. For the standardized test set, $k=200$ was used. For the challenge dataset, $k=400$ was used. This value could be fine tuned to improve the results.

Every GCN Layer uses $tanh(\cdot)$ as an activation function. All the other layers uses $relu(\cdot)$, except for the last one which uses $sigmoid(\cdot)$.

The model was trained using \emph{binary cross-entropy} as a loss function and \emph{NAdam} as the optimizer. The learning rate used was $1E-5$ with a batch Size of $1$.

The solution was implemented in TensorFlow, using Keras and the Spektral library \cite{grattarola2020graph}. 

The implementation can be found in: \url{https://github.com/franciscoandrades}

%2 etapas:

%- seleccionar datos de entrenamiento.
%* como seleccionar conjunto de entrenamiento, clase positiva y clase negativa. Que restriccion se aplico para los link de entrenamiento (solo links que tuvieran al menos uno de sus nodos con degree > 0). Como se construyeron los subgrafos. que años se eligieron para entrenar/validar. 

%* Sobre conjunto de test: que se hizo con los vertex pairs con los dos nodos con degree 0. (no se predijo sobre estos, se les asignó 0 directamente), probé con asignarles la probabilidad (obtenida empiricamente) que tiene un link con los dos nodos con degree 0 de tener target 1, pero era un numero muy cercano a 0, no cambiaba nada. 

%- SEAL con dgcnn como arquitectura.

%seal en grafo estatico utilizando arquitectura DGCNN, ignorando informacion temporal. usando solo un porcentaje de los nodos de cada subgrafo. K fue elegido de tal manera que ...

%Before you begin to format your paper, first write and save the content as a 
%separate text file. Complete all content and organizational editing before 
%formatting. Please note sections \ref{AA}--\ref{SCM} below for more information on proofreading, spelling and grammar.

%Keep your text and graphic files separate until after the text has been 
%formatted and styled. Do not number text heads---{\LaTeX} will do that 
%for you.

\section{Experiments and Results}\label{sec:results}

The proposed method is compared with the baseline described in Section \ref{sec:background} using the AUC score on 3 different scenarios: 1) Standardized test set, 2) Only Type 1 links of the standardized test set (Test$^\dagger$) and 3) Challenge Leaderboard. The experimental results are summarized in Table 2.

\begin{table}[h]
\begin{center}
\begin{tabular}{||c c c c||} 
 \cline{2-4}
\multicolumn{1}{c}{} & \multicolumn{3}{|c||}{AUC} \\
 \hline
   & Test & Test$^\dagger$  & Challenge  \\ [0.5ex] 
 \hline\hline
 MK's Baseline & 0.732 & 0.654 & 0.879  \\ 
 \hline
 This work & 0.805 & 0.775 & 0.877  \\
 \hline
\end{tabular}
\medskip
\caption{Comparisons of our method and MK's Baseline on 3 scenarios: 1) Test: Standardized test set, 2) Test$^\dagger$: Only Type 1 links on standardized test set and 3) Challenge: Challenge Leaderboard}
\end{center}
\end{table}

We can see that the method presented in this work is competitive against MK's Baseline. In our opinion, this is an interesting result as the proposed approach does not explicitly use temporal features or a model (e.g. RNN) devised to precisely learn the dynamics of the graph in the temporal dimension. Moreover, for computational reasons, the method employs only \textit{1 hop enclosing sub-graphs} and $35\%$ of the nodes enclosing a node pair, resulting in a low memory solution. 

One of the most interesting results is the AUC Score on Type 1 node pairs. These cases correspond to a kind of cold-start scenario, in which one of the nodes does not have neighbours. They may represent the situation in which an emerging concept starts to be researched by the community (positive example) or a concept that is proposed in the literature but fails to be accepted (negative example). As expected, both the proposed method and the baseline reduce their accuracy on these asymmetric cold-start cases. However, our method looks much more robust in this scenario, increasing its advantage over the baseline. This ability is useful in a range of applications.

We hypothesize that the model is learning to solve two different problems, related to the different types of node pairs it can receive as input (please see \ref{sec:background}). Type 0 examples corresponds indeed to a regular link prediction problem in which we have 2 relatively large sub-graphs around the target nodes.  We refer to this prediction as an \emph{union} problem. A Type 1 link prediction instead, takes a large sub-graph and an isolated node as an inputs. As there is no topology information about the isolated node, it can be interpreted as irrelevant to the output. Therefore, in this case, the task of link prediction can be formulated as predicting the probability that has the sub-graph of absorbing a new, arbitrary node in a particular position of its topology. We refer to this prediction as a \emph{node absorption} problem.

%The problems of link prediction and node absorption are fundamentally different. We speculate that this two problems are solved by different learnable functions. 

To explore this hypothesis, we study whether the activations of our trained network show evidence of two different learned functions. Our experiments where made on the output of the SortPooling layer as it has a fixed size, consistent order and captures the totality of the latent representations computed by the GCN layers. The 100 best predicted training examples for each label where used for the forward pass to extract the activations. Then, five of them were randomly selected for visualization.

\begin{figure}
    \centering
    
    \caption{SortPooling activations of 5 training examples of Type $0$ with label $1$}
    \label{fig:my_label}
%\begin{tabular}{c}
 \includegraphics[scale=0.28]{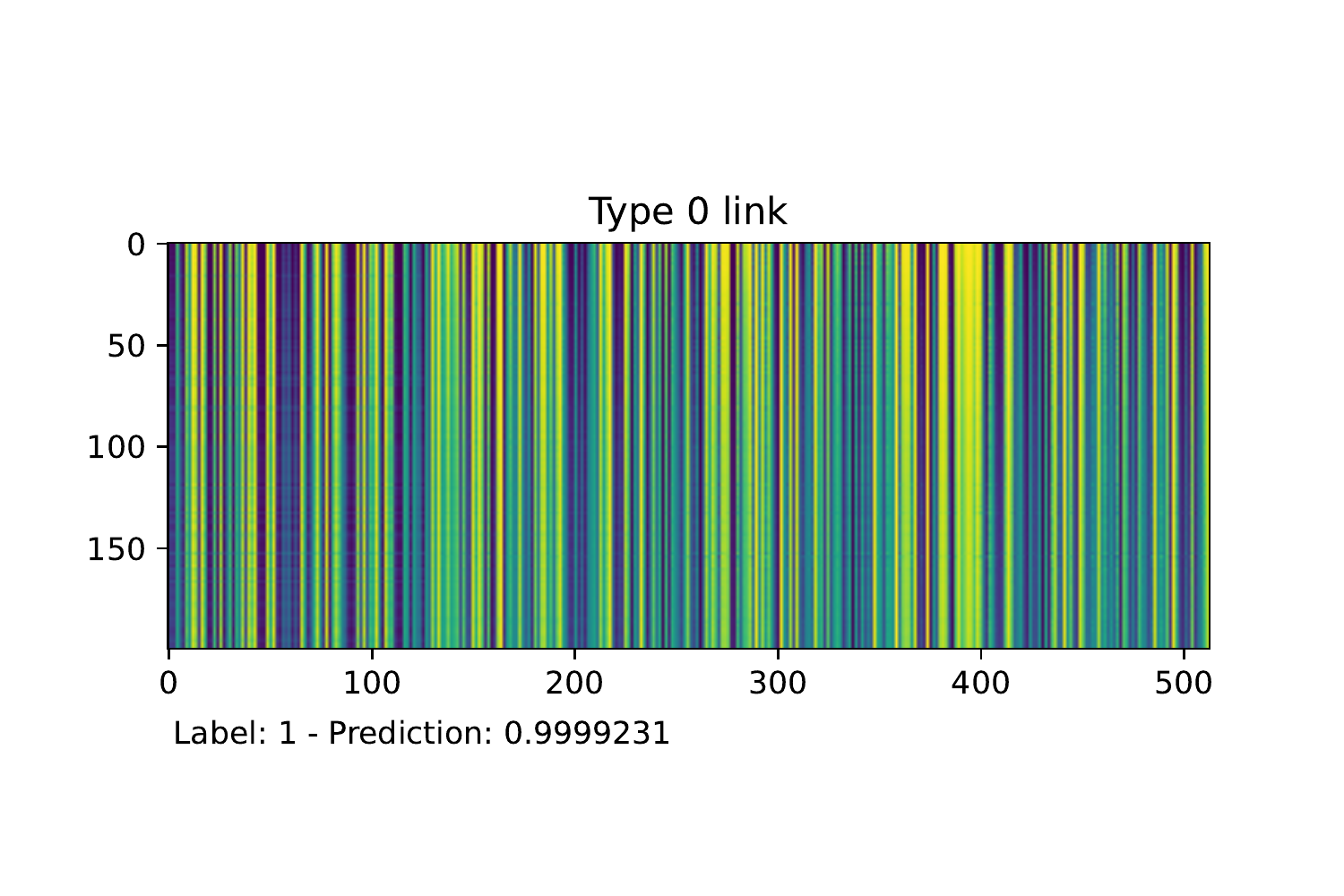}   \includegraphics[scale=0.28]{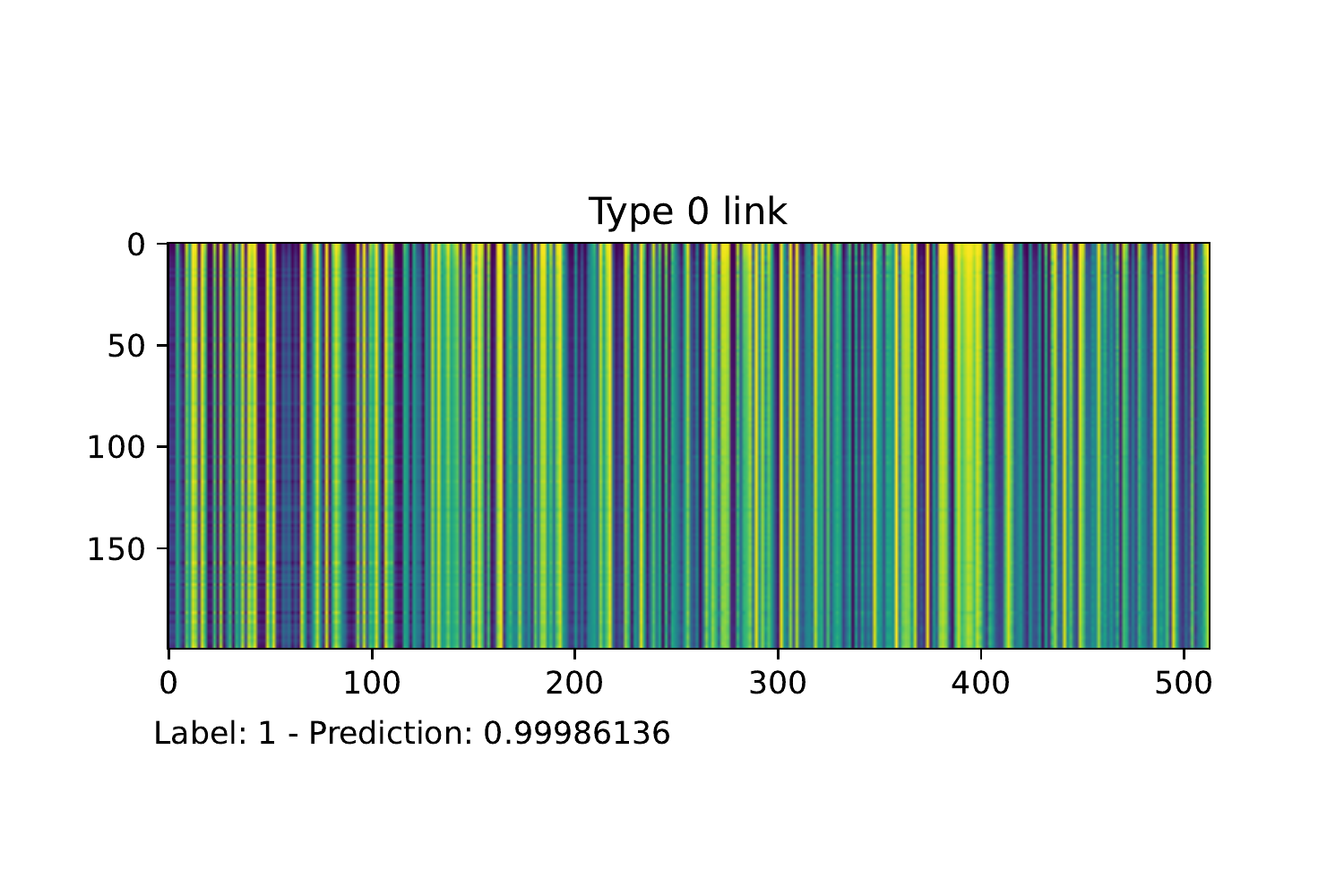}
  \includegraphics[scale=0.28]{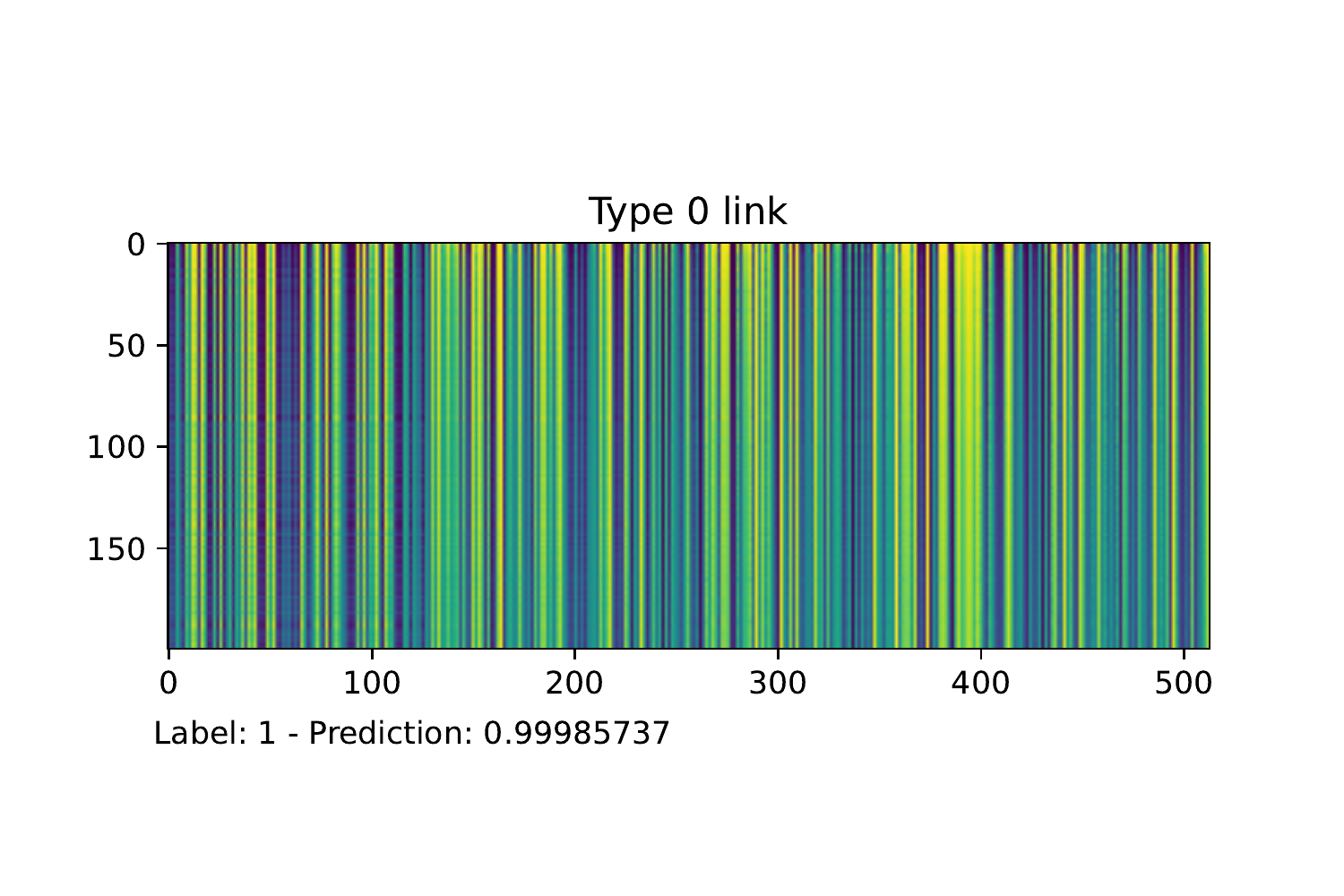}
  \includegraphics[scale=0.28]{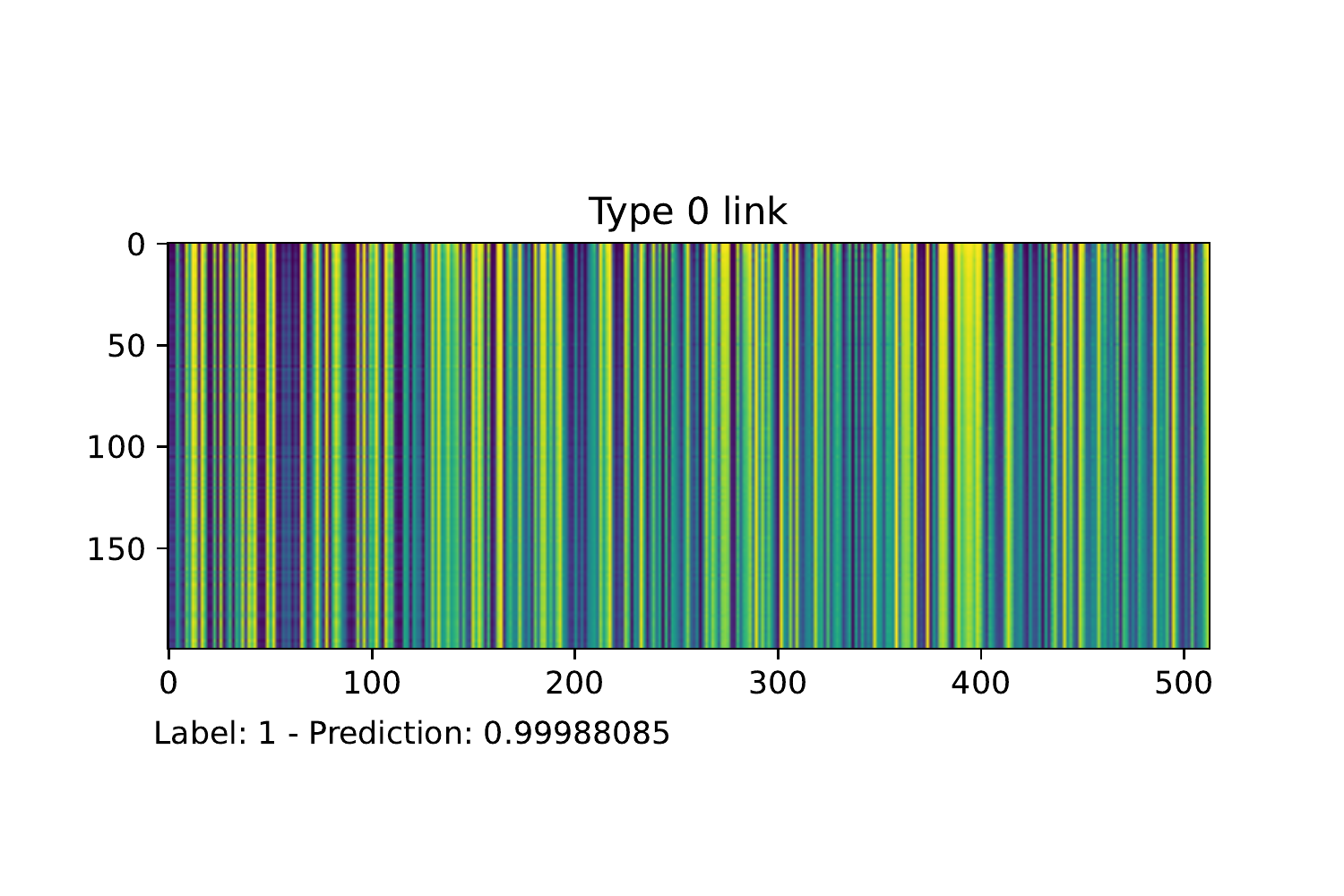}
  \includegraphics[scale=0.28]{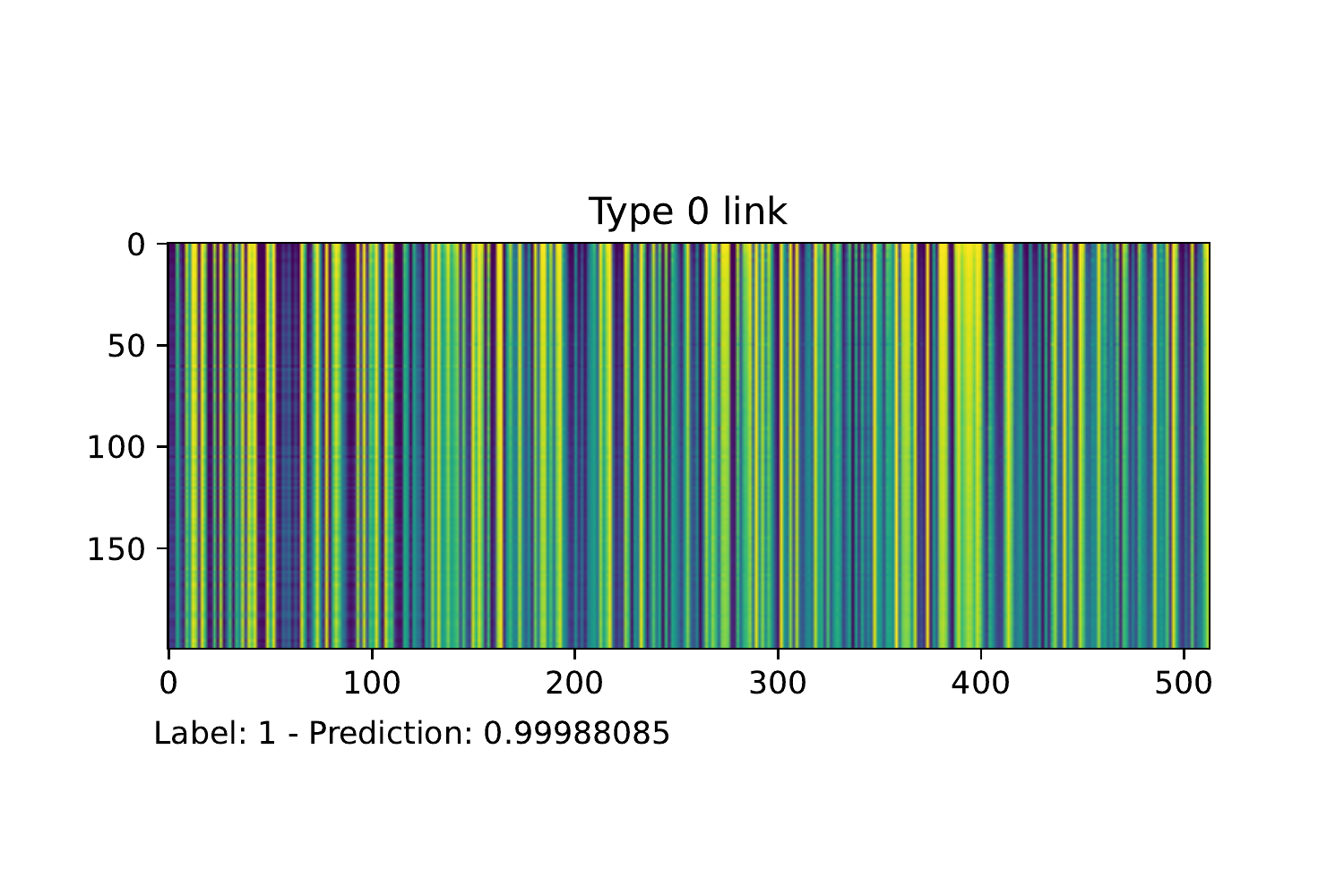}
%\end{tabular}
\end{figure}

\begin{figure}
    \centering
    
    \caption{SortPooling activations of 5 training examples of Type $1$ with label $1$ (randomly chosen).}
    \label{fig:my_label2}

\includegraphics[scale=0.28]{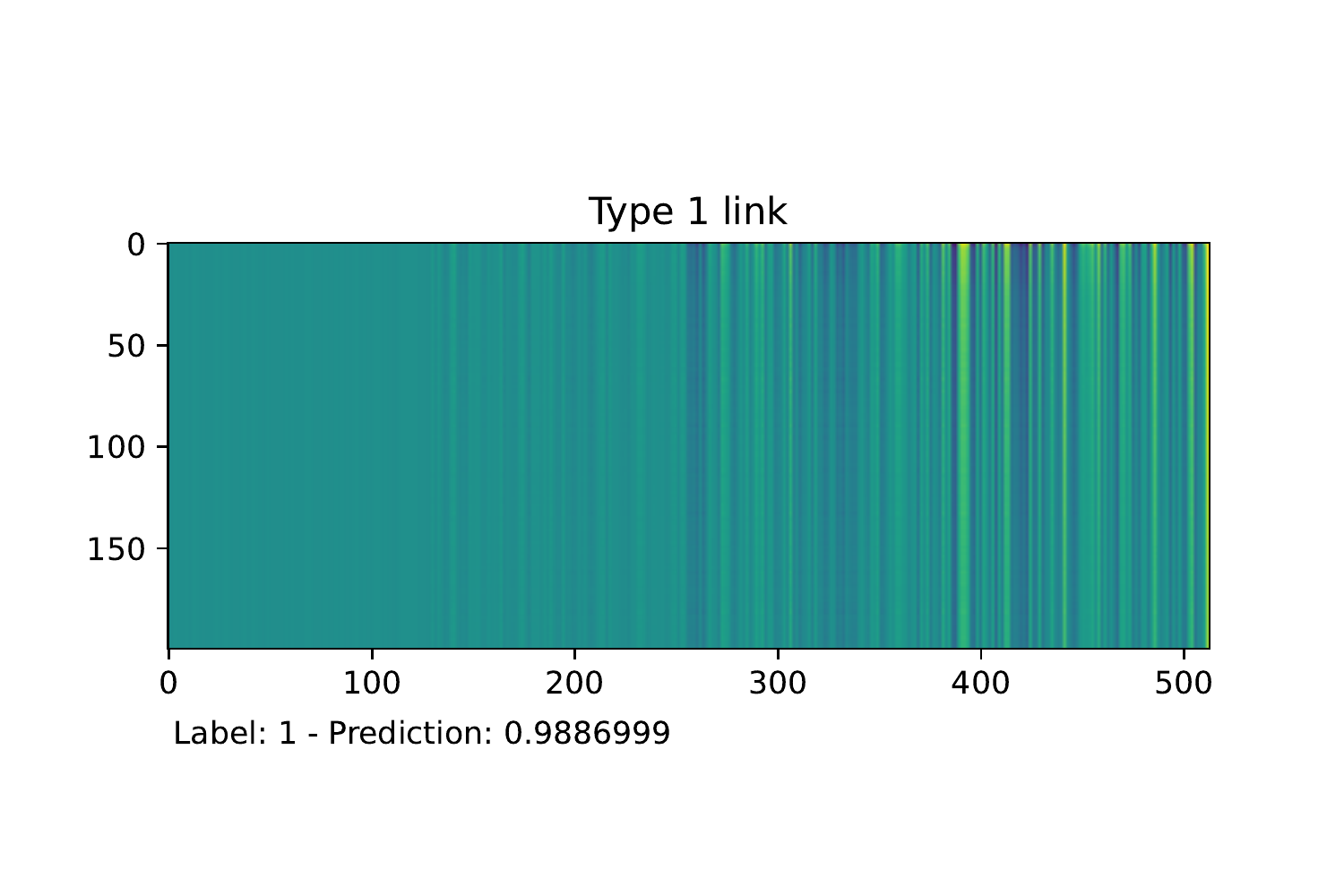}
\includegraphics[scale=0.28]{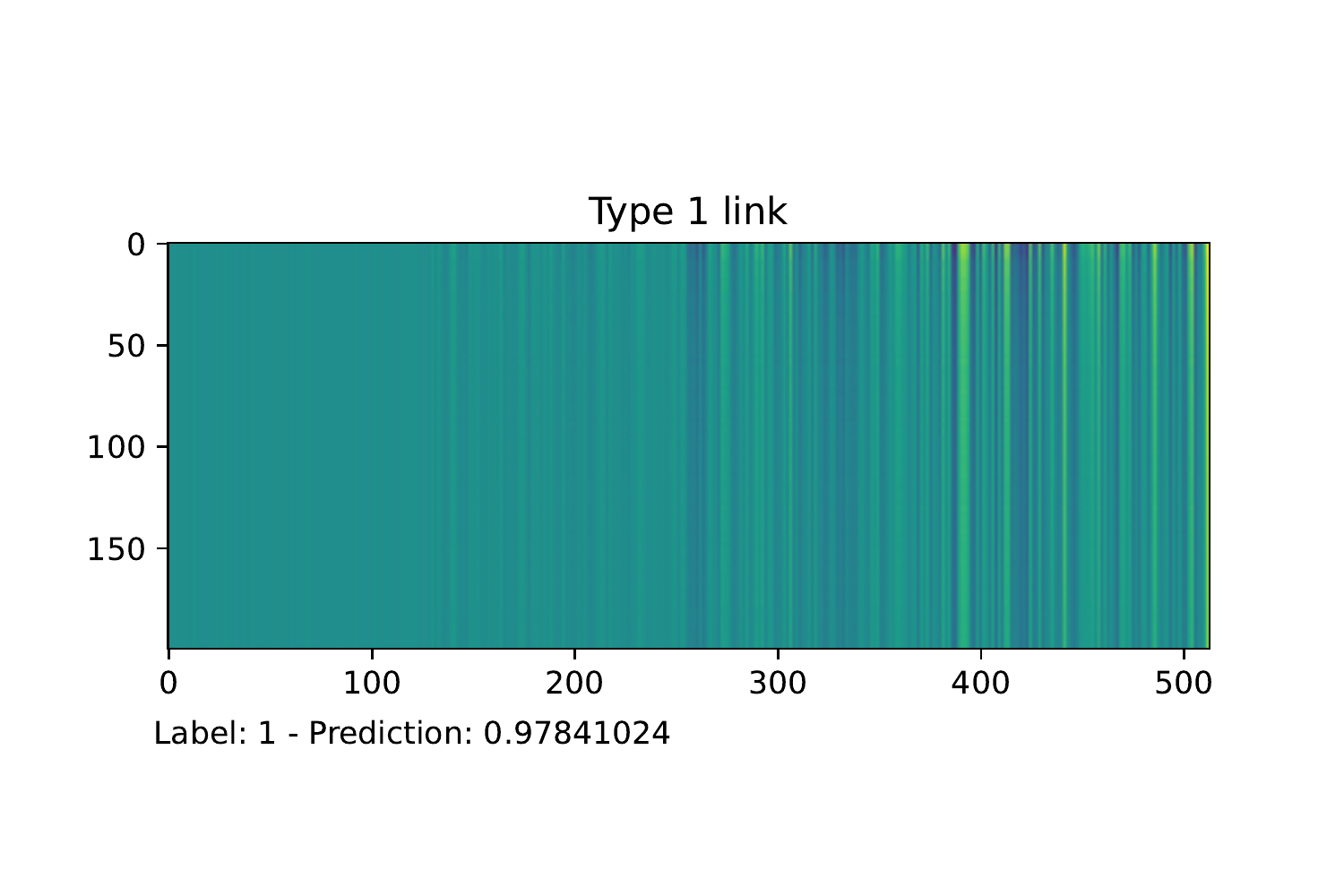}
\includegraphics[scale=0.28]{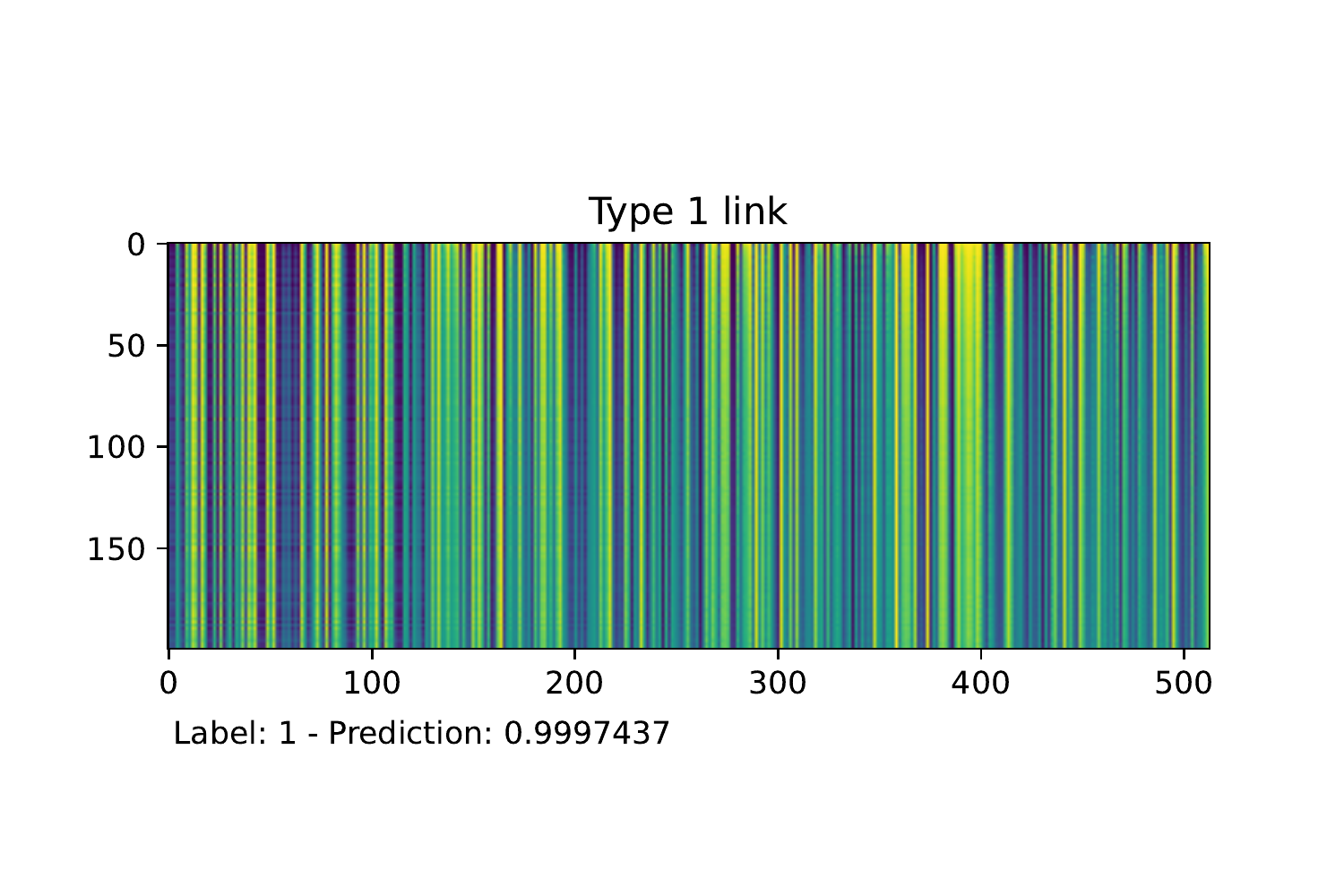}
\includegraphics[scale=0.28]{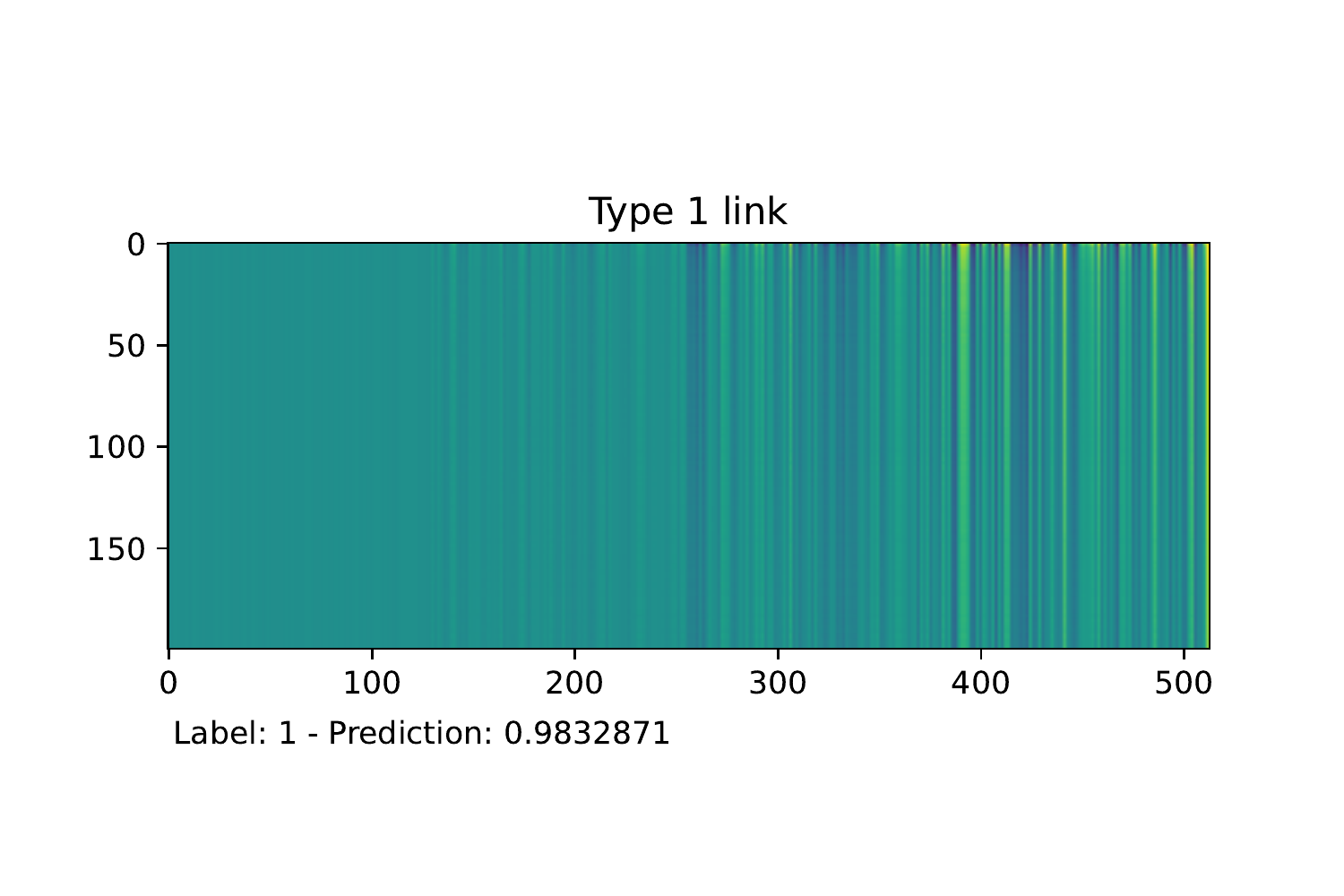}
\includegraphics[scale=0.28]{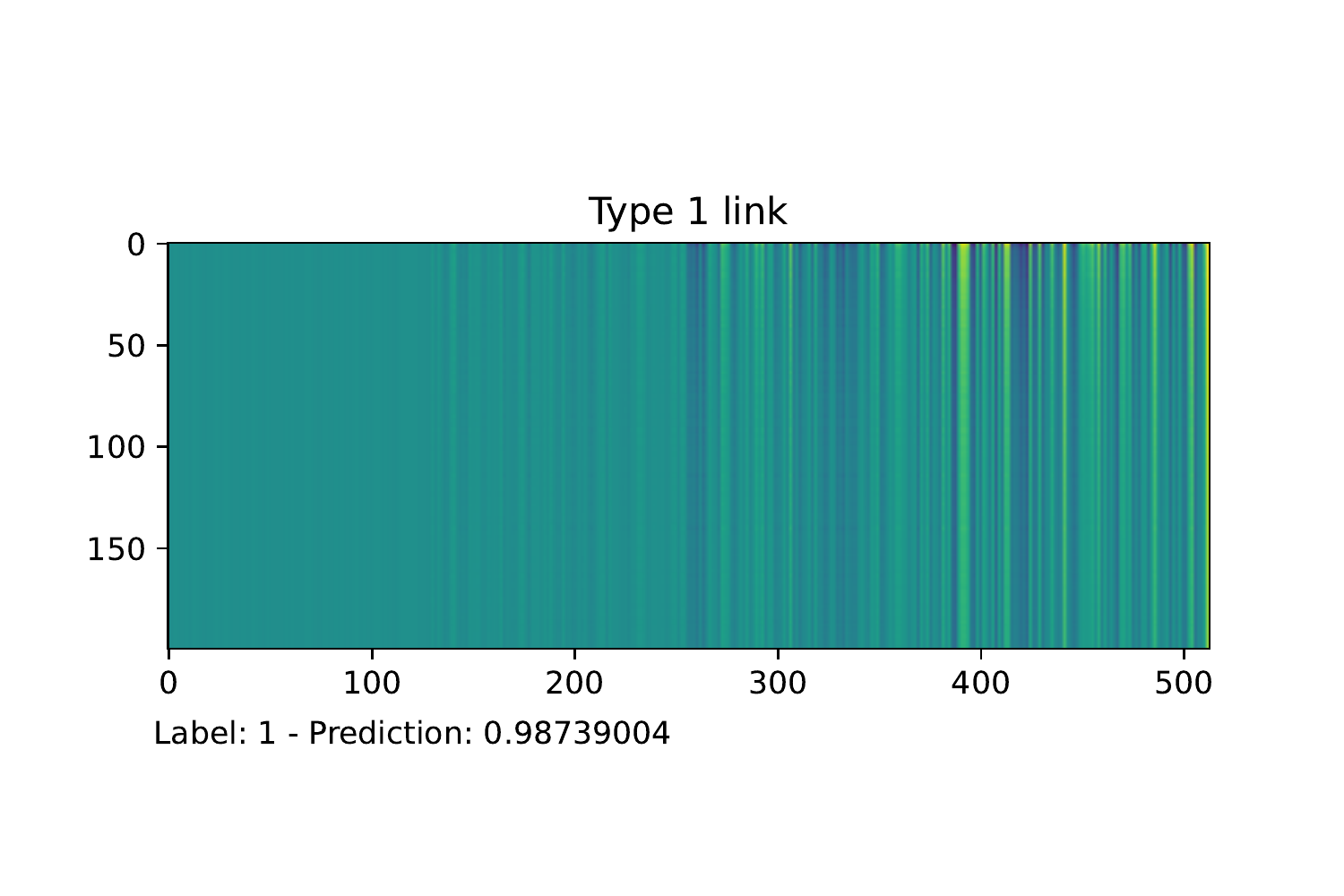}
\end{figure}

\begin{figure}
    \centering
    
    \caption{SortPooling activations of 5 training examples of Type $0$ with label $0$ (randomly chosen).}
    \label{fig:my_label3}
\includegraphics[scale=0.28]{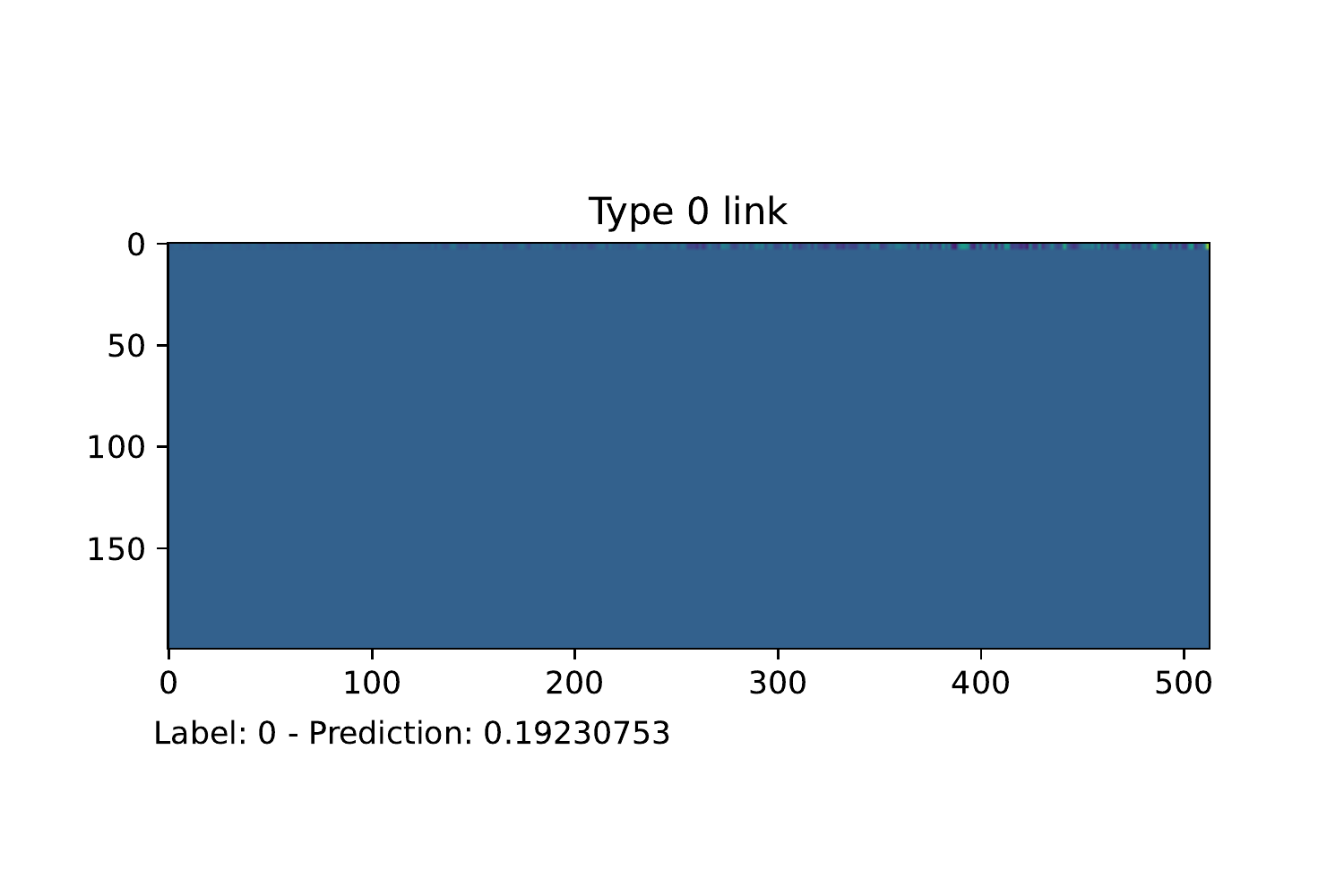}
\includegraphics[scale=0.28]{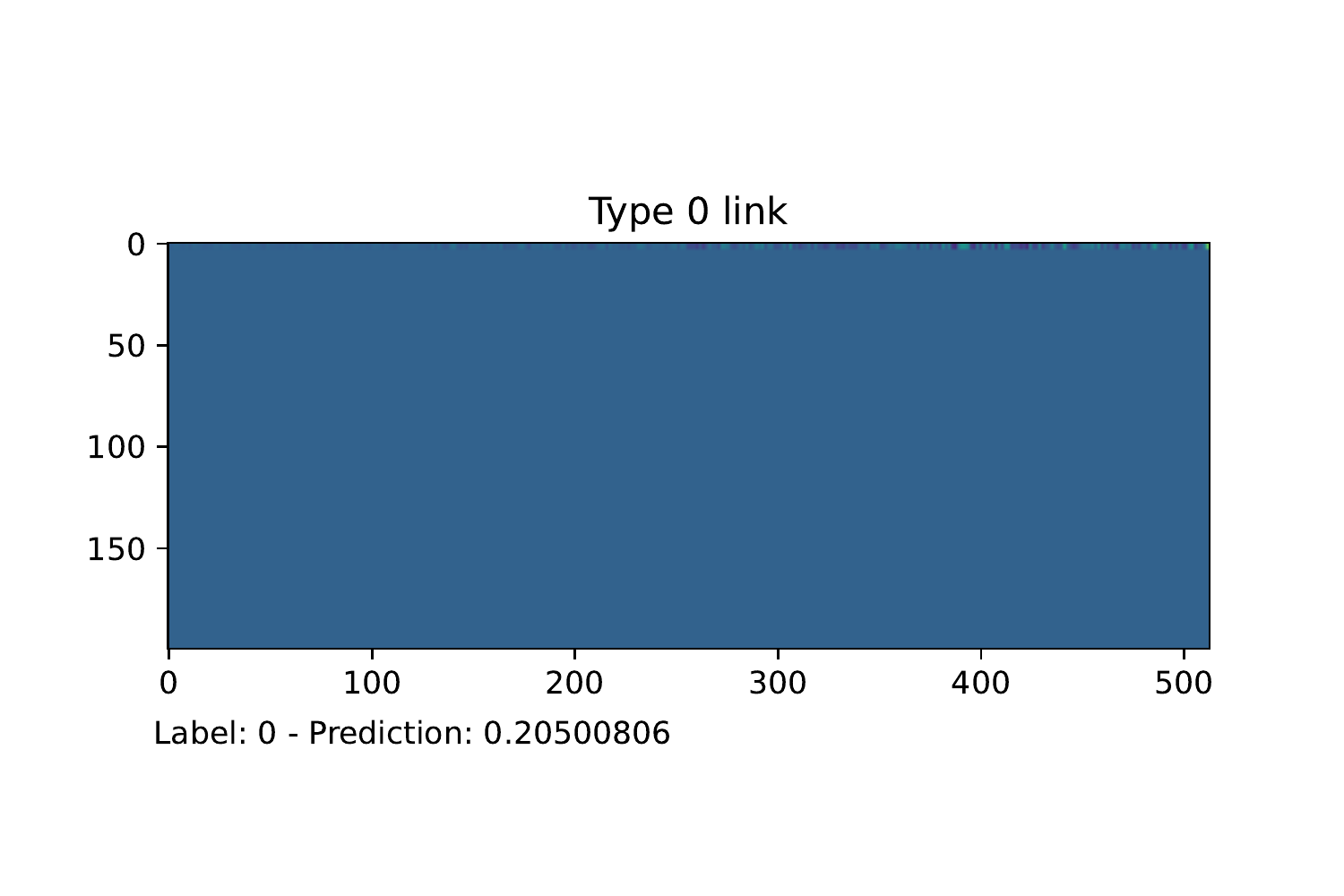}
\includegraphics[scale=0.28]{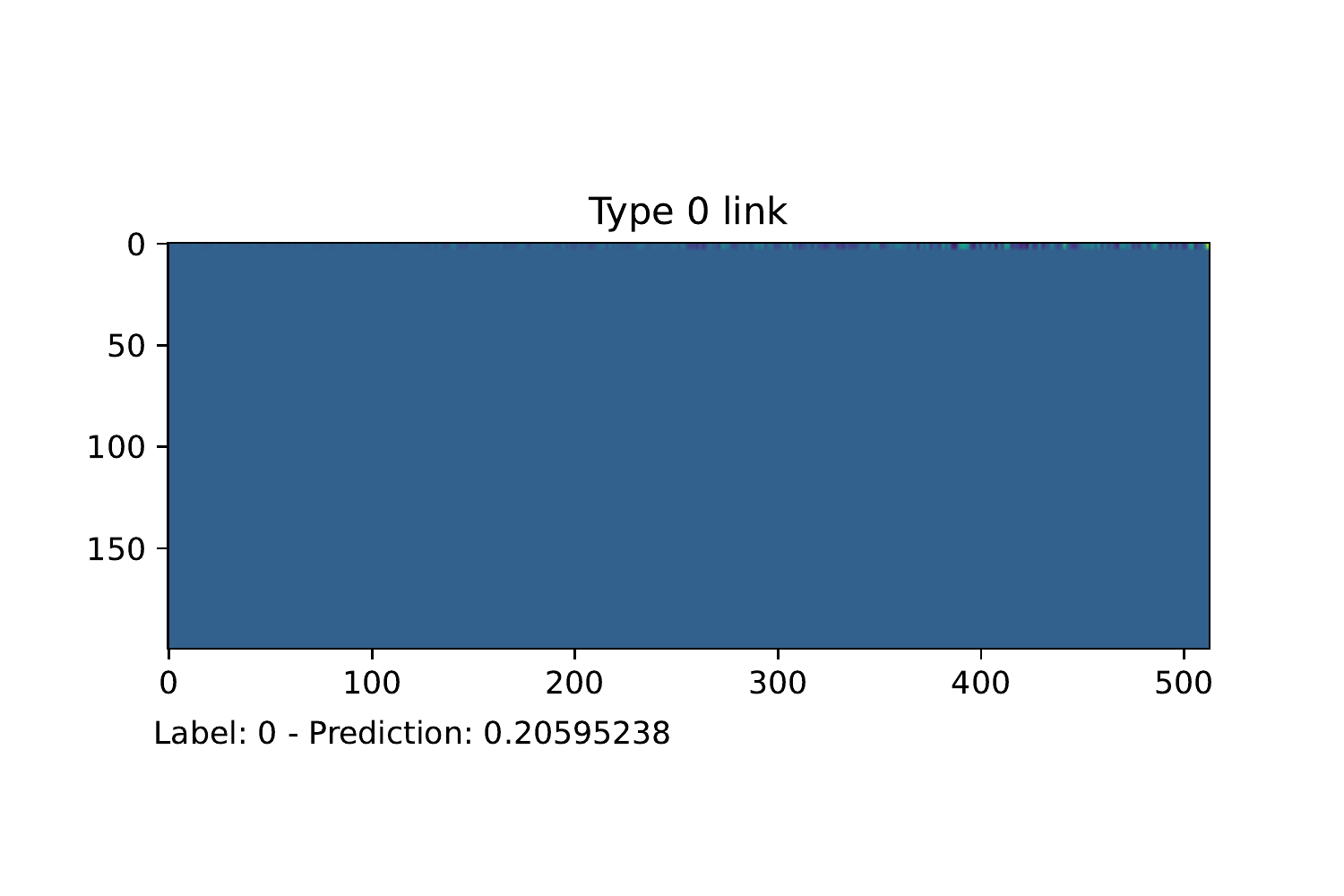}
\includegraphics[scale=0.28]{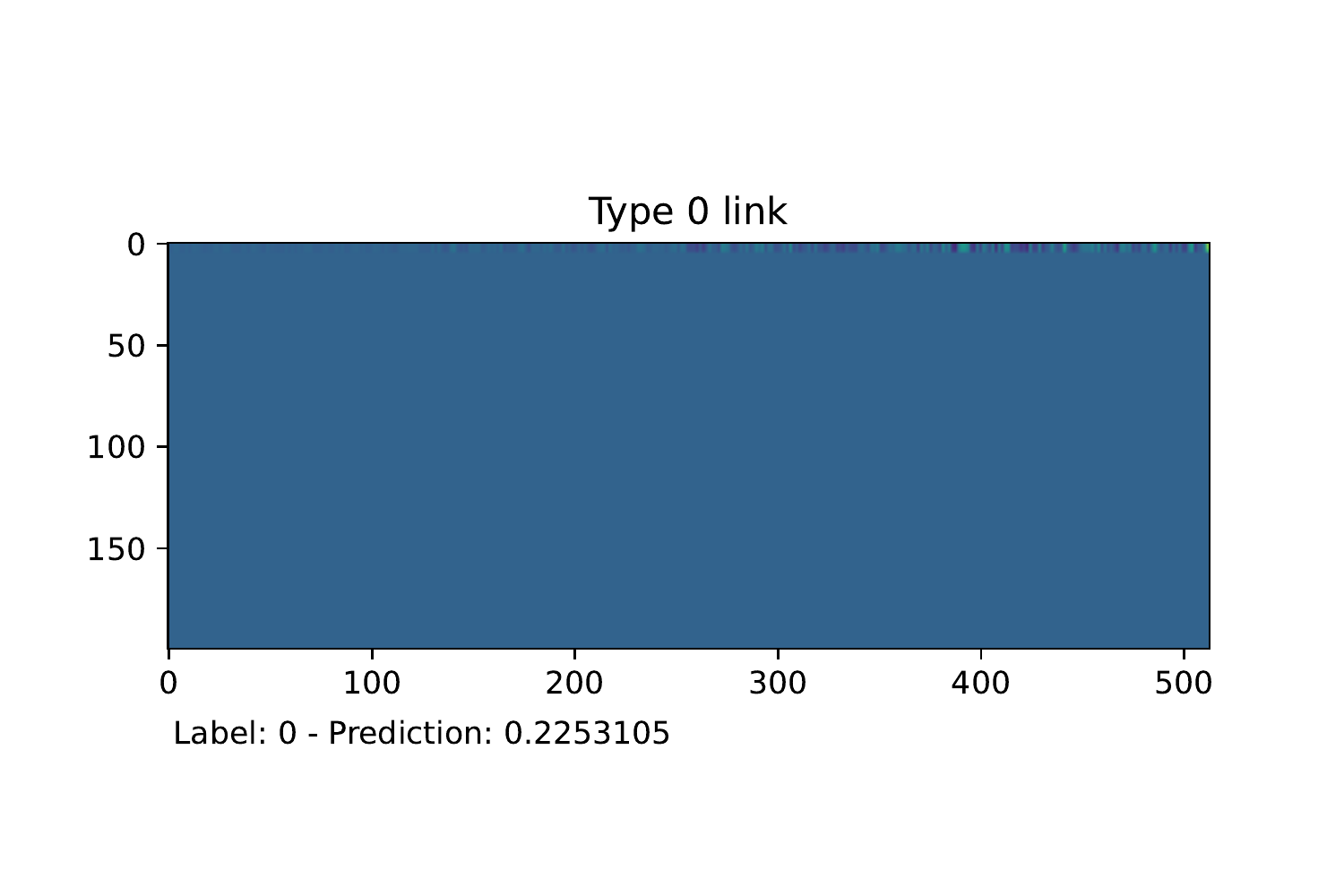}
\includegraphics[scale=0.28]{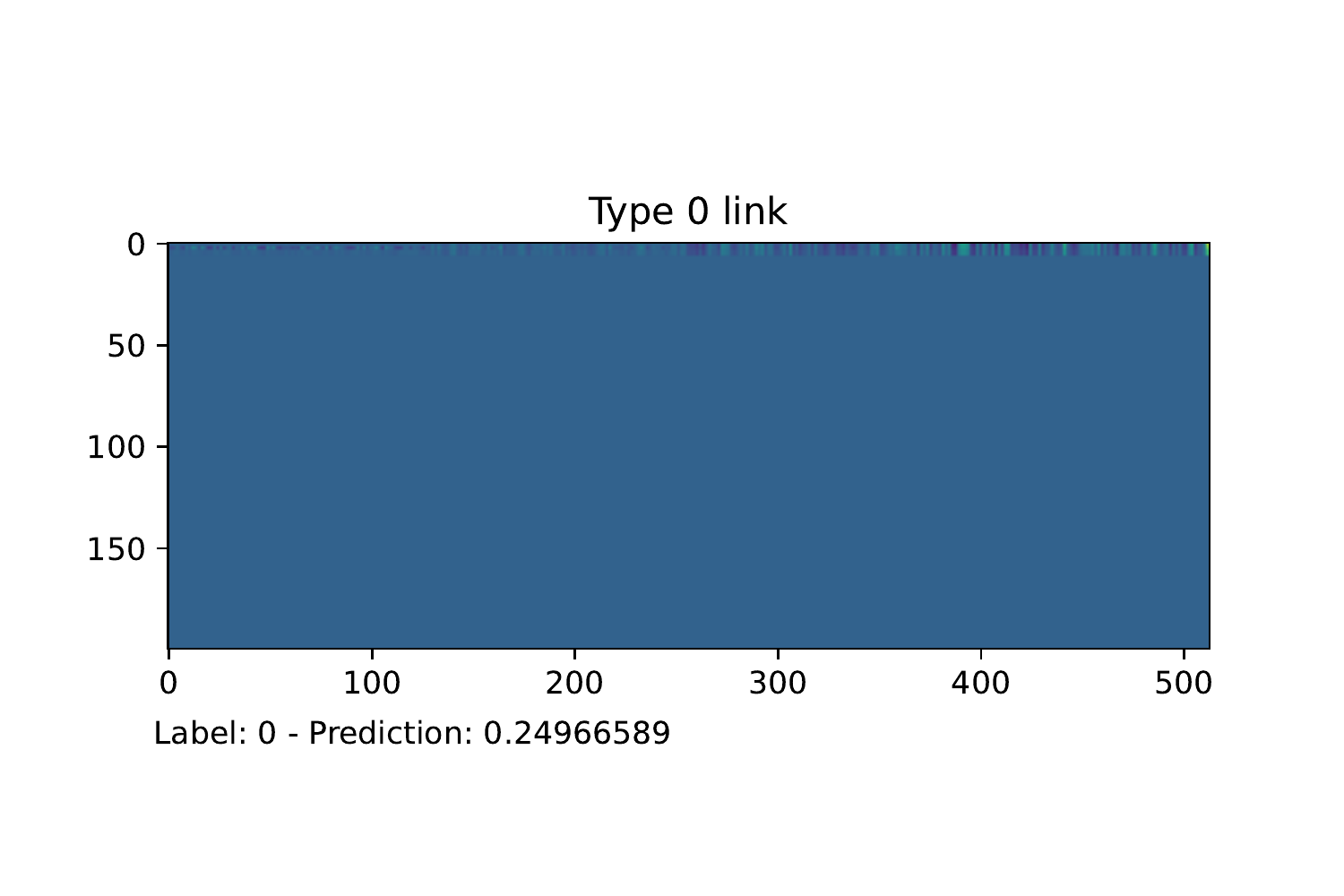}
\end{figure}

\begin{figure}
    \centering
    
    \caption{SortPooling activations of 5 training examples of Type $1$ with label $0$ (randomly chosen).}
    \label{fig:my_label4}

\includegraphics[scale=0.28]{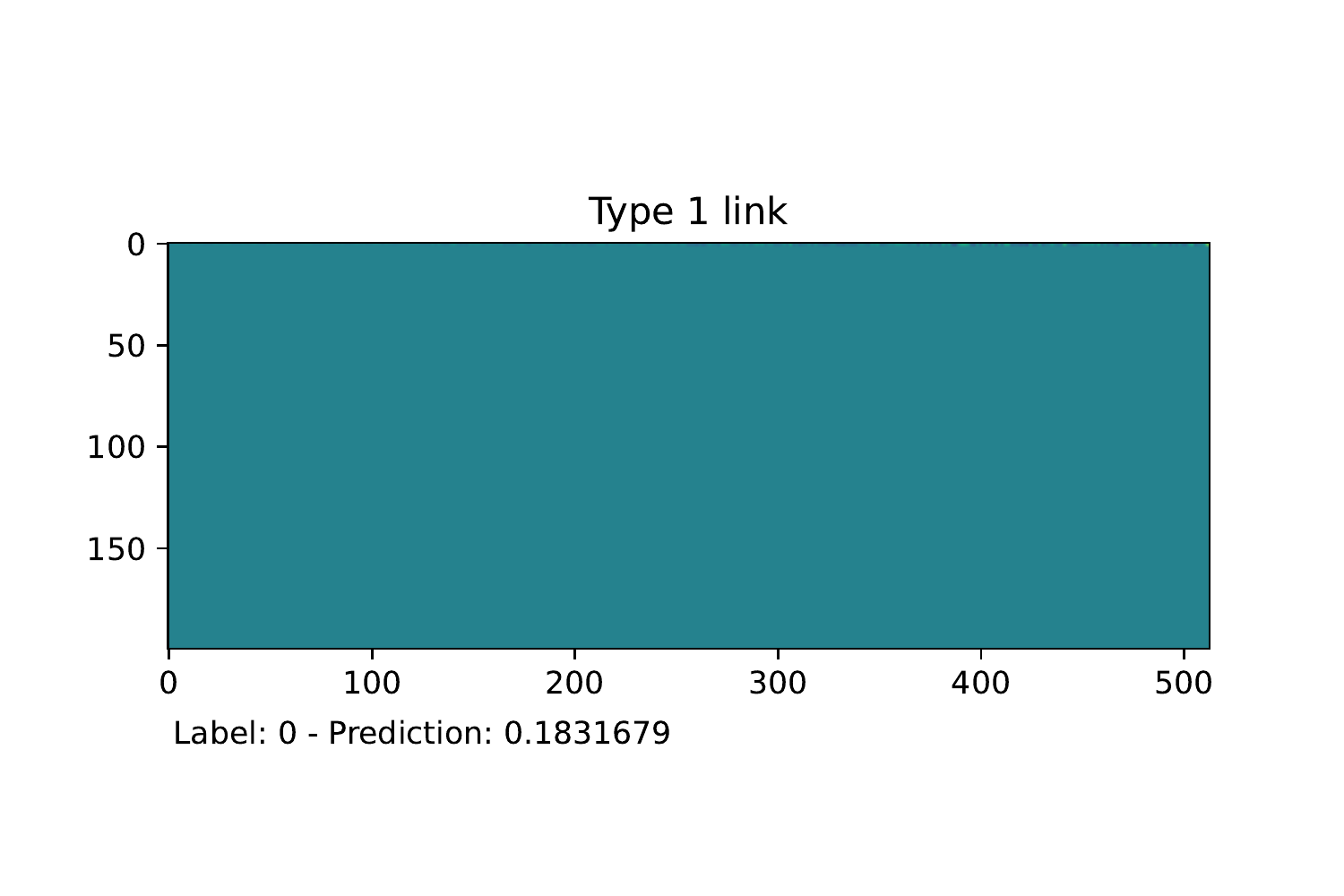}
\includegraphics[scale=0.28]{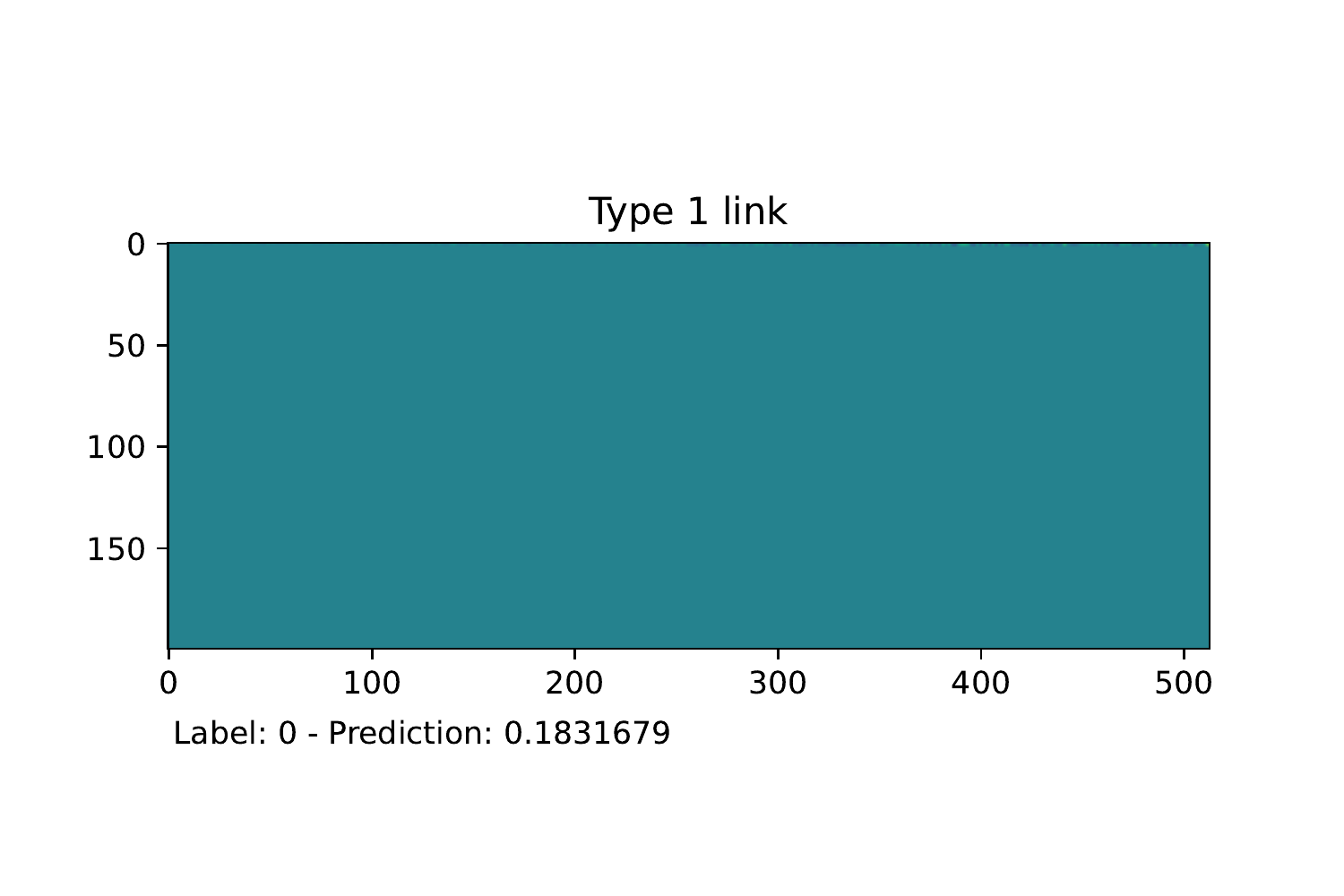}
\includegraphics[scale=0.28]{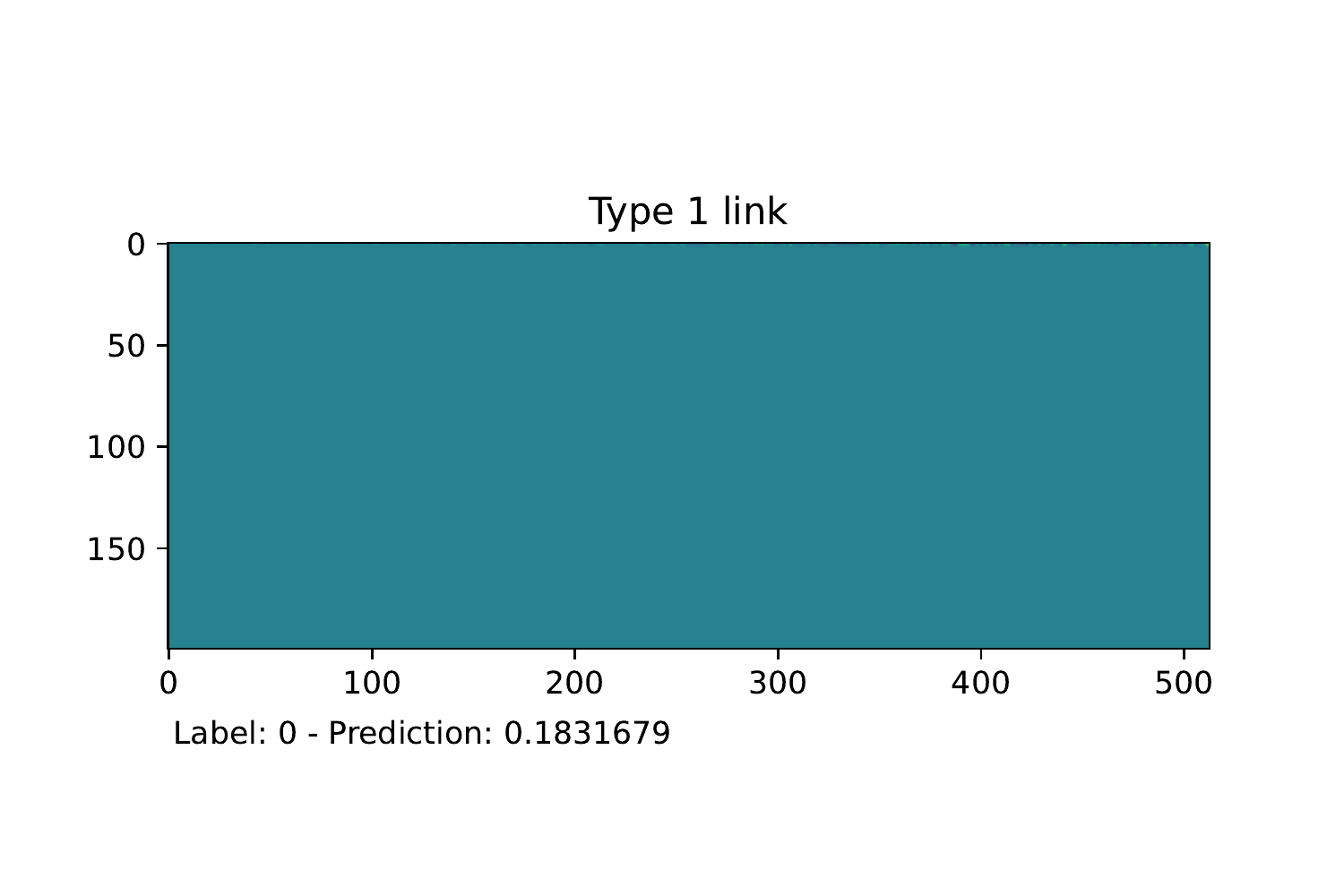}
\includegraphics[scale=0.28]{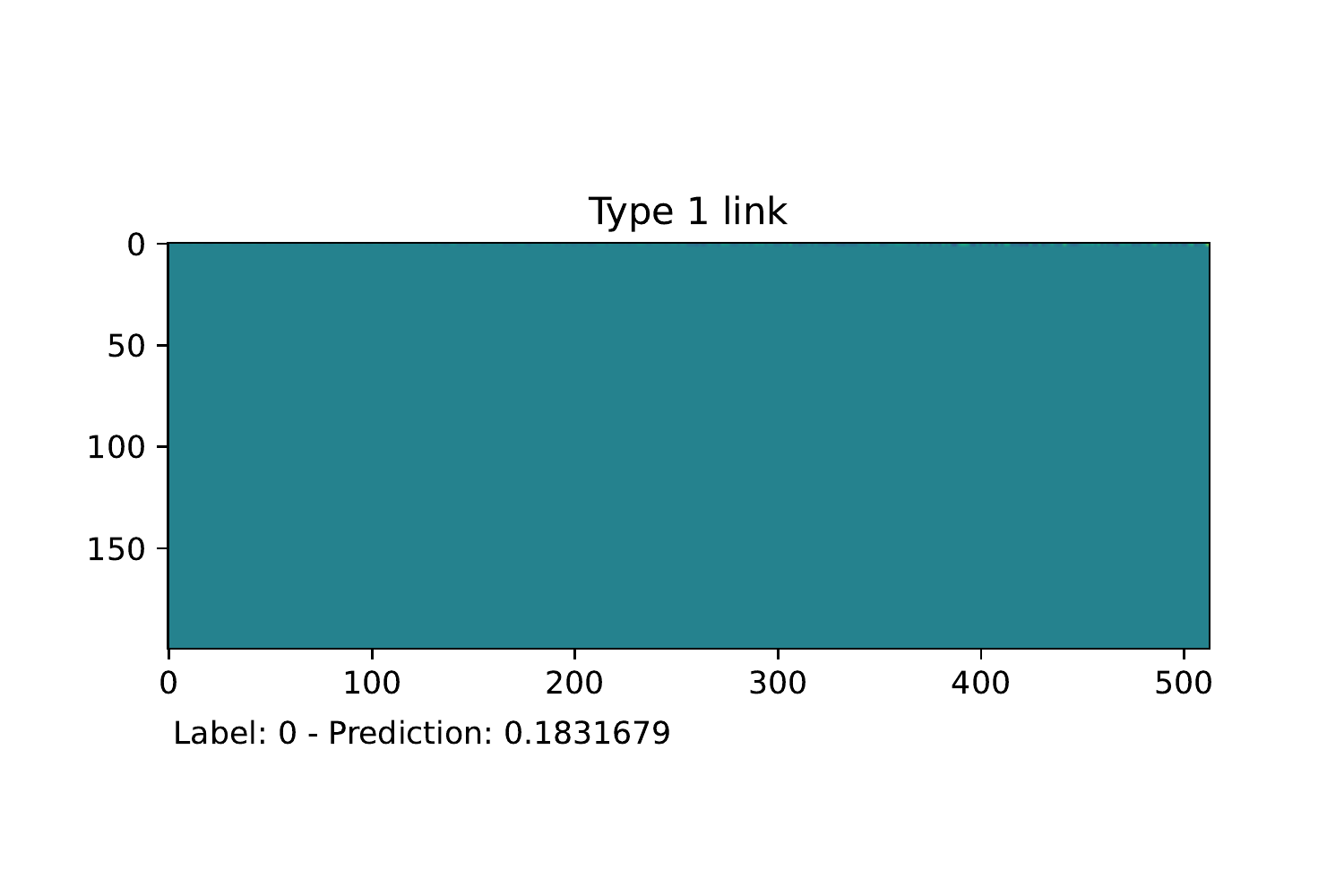}
\includegraphics[scale=0.28]{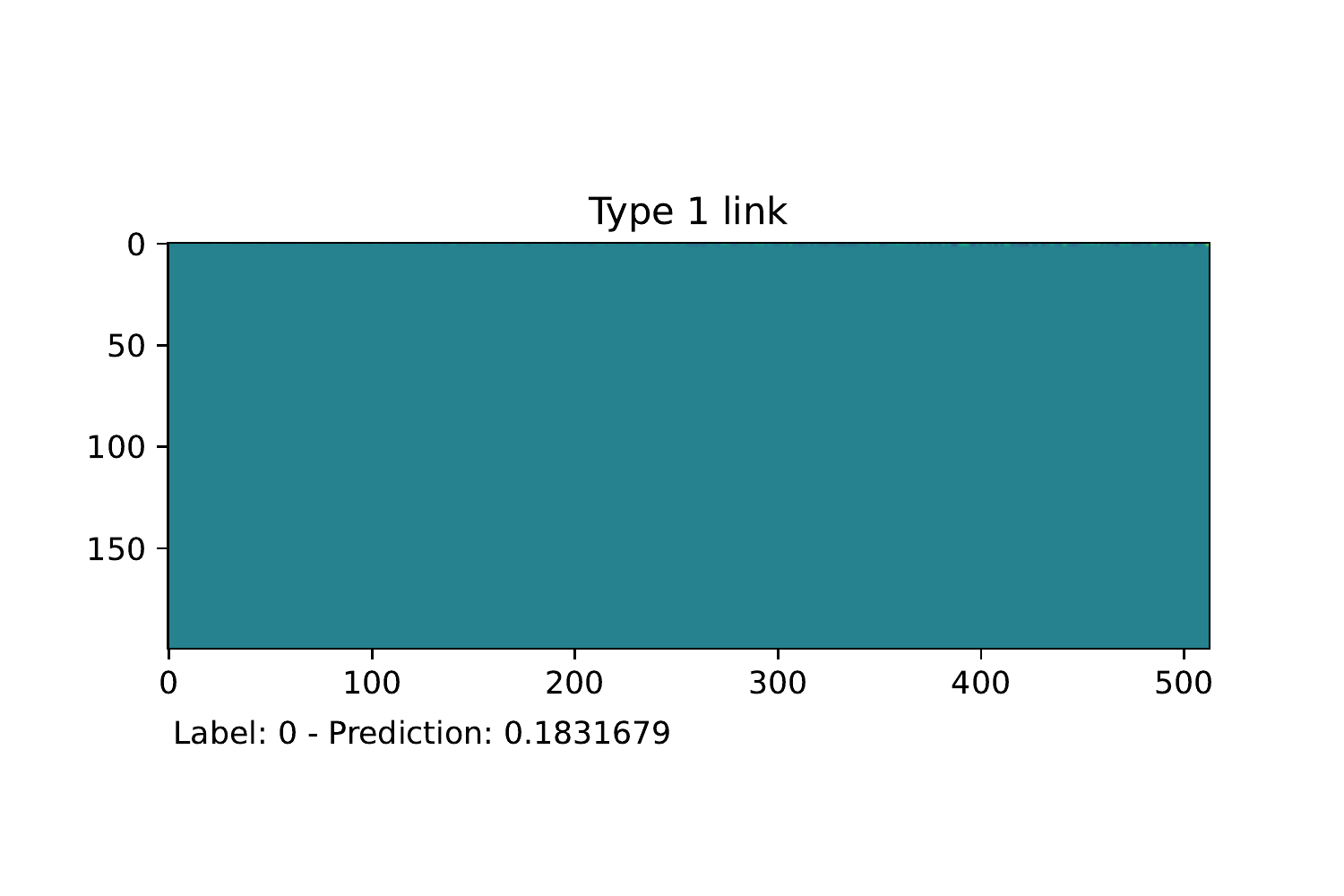}
\end{figure}

Figures~\ref{fig:my_label} and \ref{fig:my_label2} show the activations triggered by Type 0 and Type 1 positive examples (label 1). Figures~\ref{fig:my_label3} and \ref{fig:my_label4} show the activations triggered by Type 0 and Type 1 negative examples (label 0). It appears that the model indeed produces similar activations on positive examples of the same type of link prediction problem (\emph{union} vs \emph{node absorption}) but produces very different activations on examples of a different type of link prediction problem. The same can be observed as regards negative examples of the two types of link prediction problem. 

\begin{figure}
    \centering
    \caption{t-SNE embeddings on training examples with label $1$. First row: colors assigned according to node pair Type (see Section \ref{sec:background}). Second row: colors assigned according to maximum degree of the two nodes.}
    \label{fig:my_label5}
\includegraphics[scale=0.58]{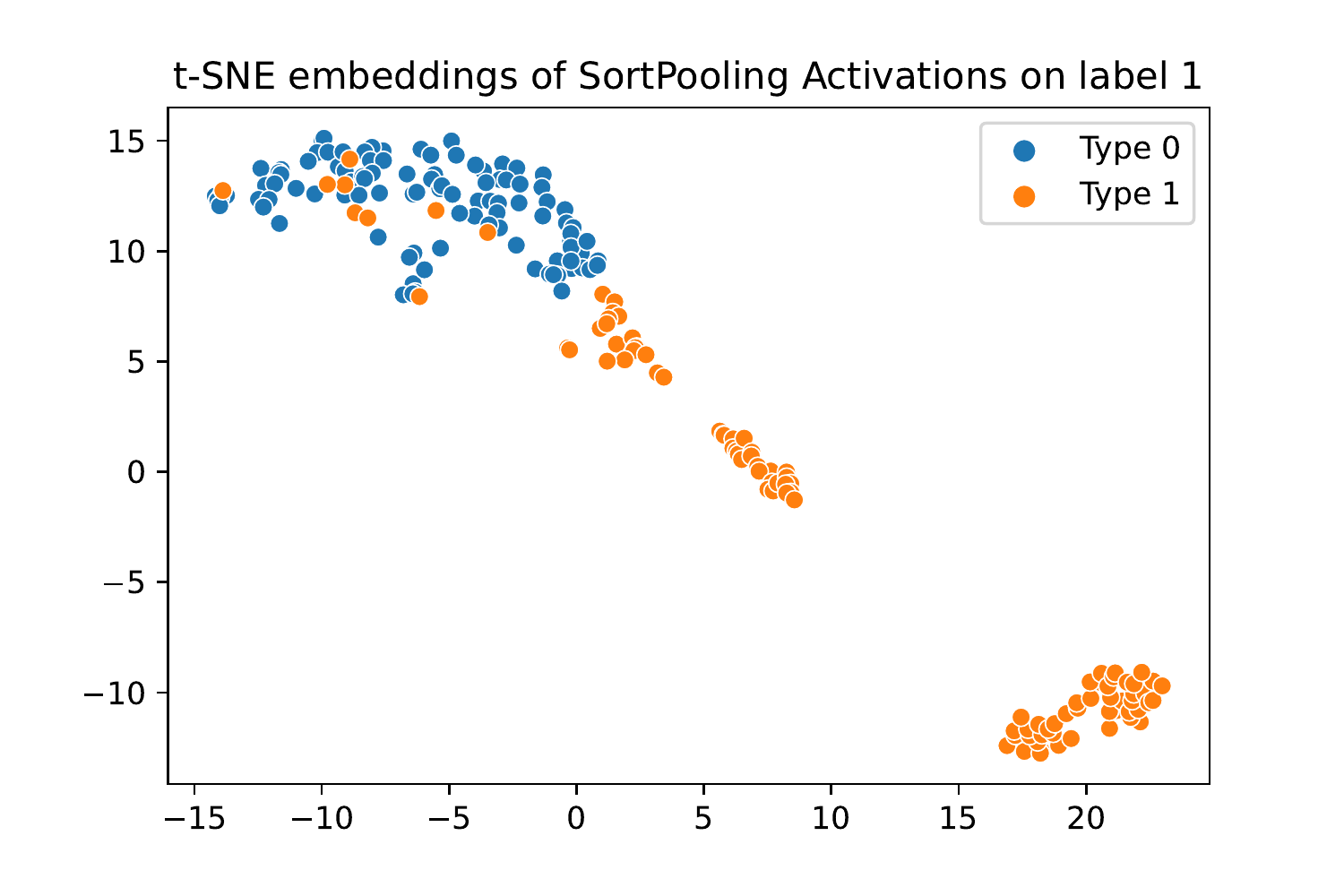}
\includegraphics[scale=0.58]{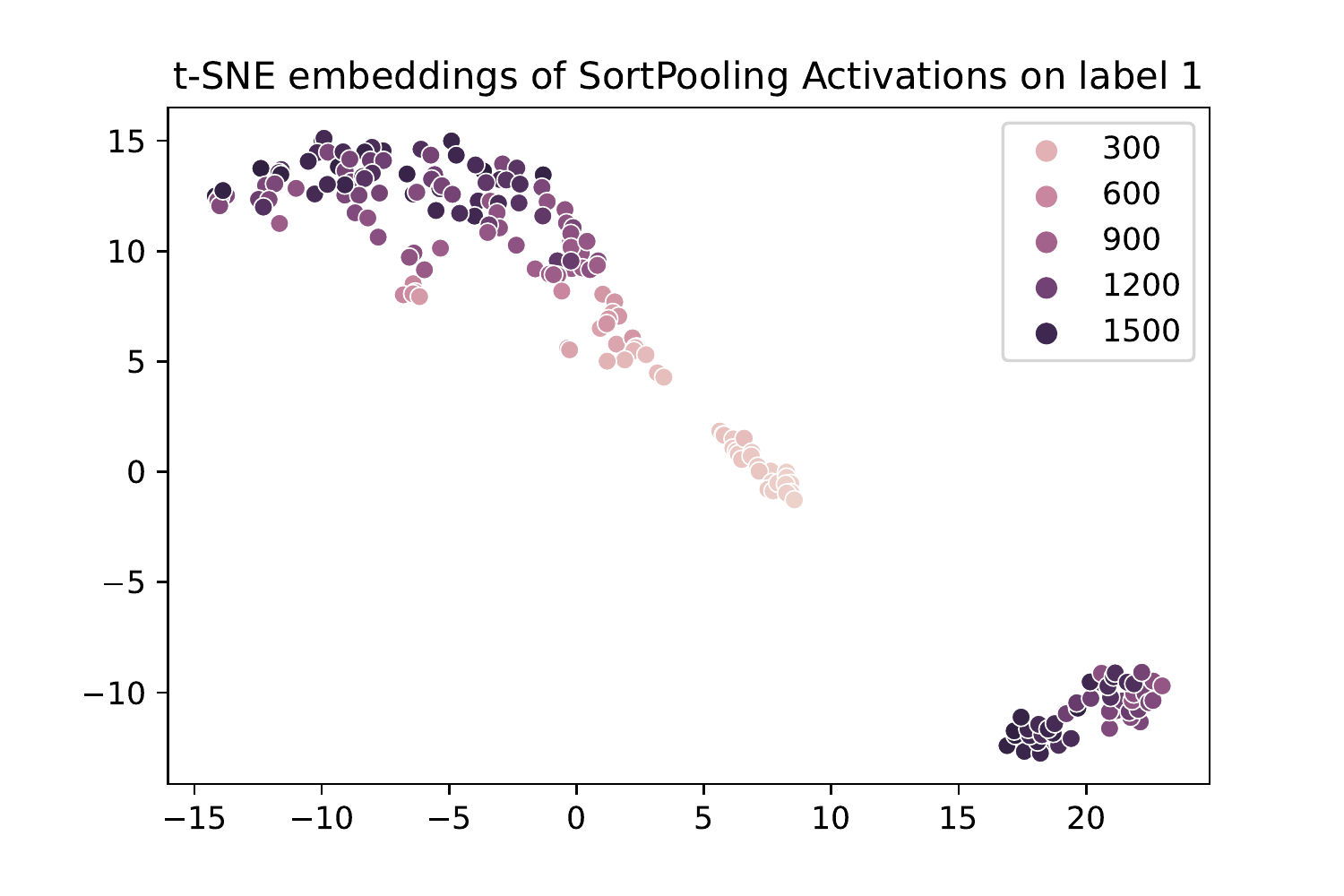}

\end{figure}

\begin{figure}
    \centering
    \caption{t-SNE embeddings on training examples with label $0$. First row: colors assigned according to node pair type (see Section \ref{sec:background}). Second row: colors assigned according to maximum degree of the two nodes.}
    \label{fig:my_label6}
\includegraphics[scale=0.58]{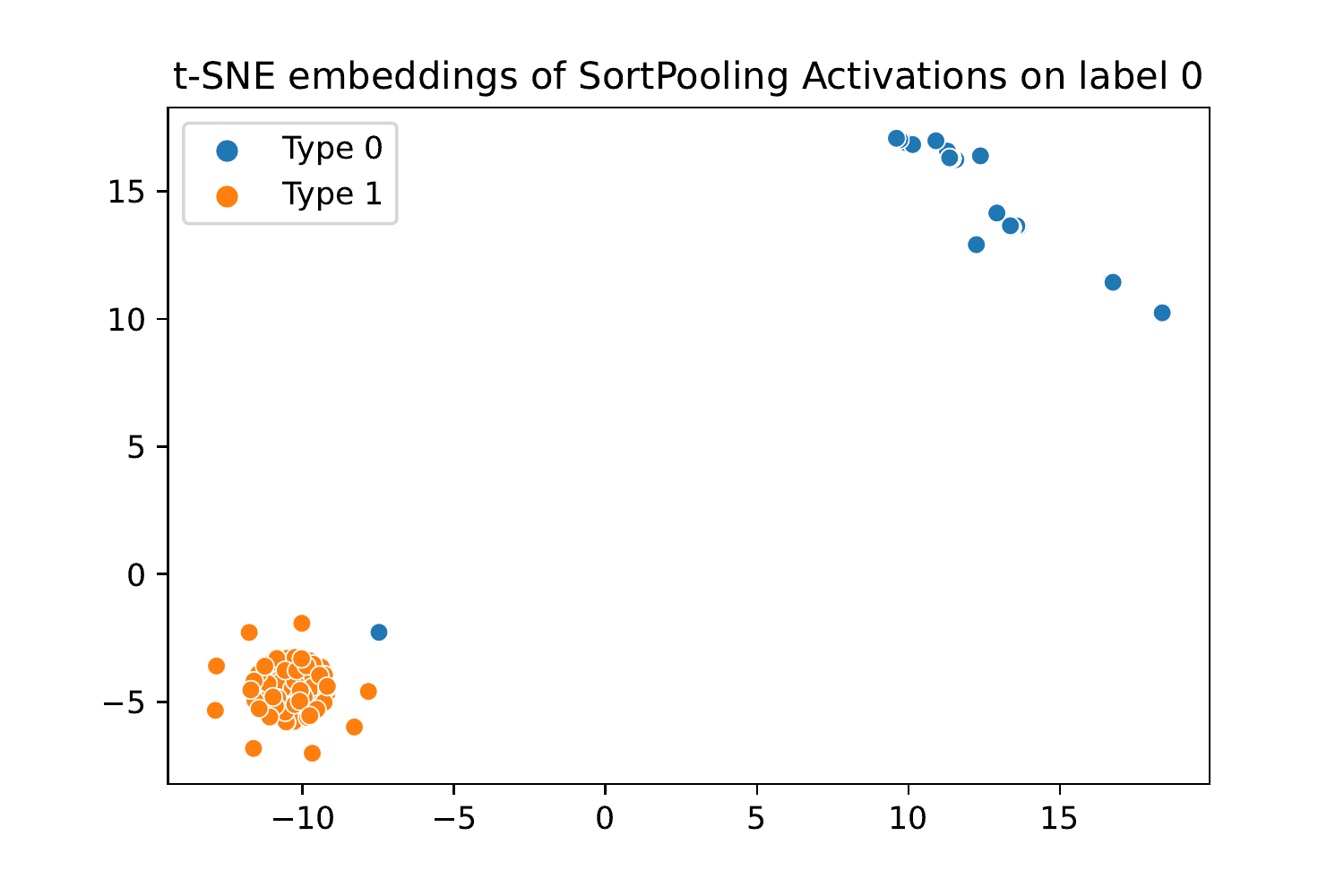}
\includegraphics[scale=0.58]{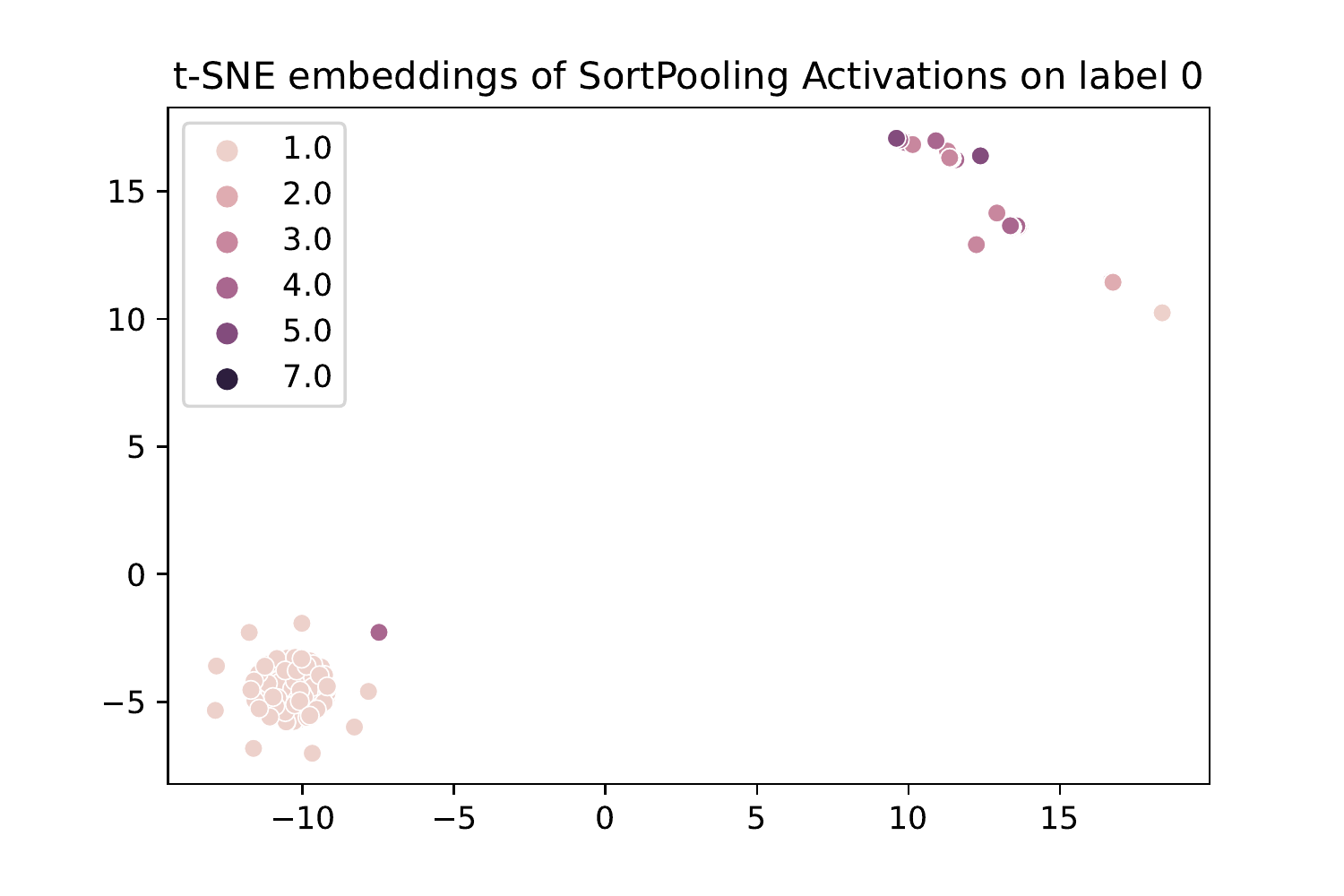}
\end{figure}

In Figures \ref{fig:my_label5} and \ref{fig:my_label6} we use t-SNE embeddings to visualize the SortPooling activations for examples with label $1$ and $0$, respectively. It appears that the examples of the same type form differentiated clusters even if they have the same label. For example, in Figure~\ref{fig:my_label6}, a Type 0 link that have both nodes with degree $1$ is still properly differentiated. This result suggests that the reason of the differentiation does not regard the complexity of the sub-graphs, but the type of the problem addressed (\emph{union} vs \emph{node absorption}). 

Altogether, results in this preliminary study suggest it is worth investigating more how the model deals with two substantially different phenomena of link emergence in semantic networks.    

% profundizar
% dejar claro que sigue siendo especulativo y hay que buscar mas evidencia

\section{Conclusions and Final Remarks}\label{sec:conclusions}
% aqui no se bien que concluir profe.

In this paper we presented a solution to the Science4cast competition. Building on recent advances in Graph Neural Networks (GNN), we approached the task as a graph classification problem. Given a node pair, we compute the sub-graphs enclosing the nodes and use stack of graph convolutional layers to predict the emergence of a link in the future. For the sake of efficiency, the model was trained by drawing input data from a single outdated state of the network (2014). Moreover, to reduce memory requirements, a small subset of nodes in the neighbour of the target link was presented to the neural net.

The results of the challenge show that despite the strong restrictions we imposed to obtain a parsimonious model, the solution is competitive and scores 0.877 AUC in the public leaderboard. In this paper, we also presented preliminary results that suggest that our solution is more robust to an asymmetric cold start situation in which one node does not have neighbours. In these cases, the solution increased its advantage over the baseline. As this scenario represents the situation in which an emerging concept starts to be researched by the community or fails to be accepted, we believe it is worth conducting more experiments (e.g. visualization strategies) that help to understand the patterns learnt by the model.

In future work, we plan to equip the model with sampling strategies which exploit the edge weights available in the dataset. We also plan to use the GNN trained for the competition to extract features on which a recurrent model could be trained.

\section*{Acknowledgment}

The first author acknowledges the Scotiabank Centre for Digital Transformation at the Federico Santa Maria University for the funding support to attend the IEEE Big Data Conference 2021.

%The preferred spelling of the word ``acknowledgment'' in America is without 
%an ``e'' after the ``g''. Avoid the stilted expression ``one of us (R. B. 
%G.) thanks $\ldots$''. Instead, try ``R. B. G. thanks$\ldots$''. Put sponsor 
%acknowledgments in the unnumbered footnote on the first page.

%\section*{References}

\bibliography{references.bib}
\bibliographystyle{IEEEtran}

%Please number citations consecutively within brackets \cite{b1}. The 
%sentence punctuation follows the bracket \cite{b2}. Refer simply to the reference 
%number, as in \cite{b3}---do not use ``Ref. \cite{b3}'' or ``reference \cite{b3}'' except at 
%the beginning of a sentence: ``Reference \cite{b3} was the first $\ldots$''

%Number footnotes separately in superscripts. Place the actual footnote at 
%the bottom of the column in which it was cited. Do not put footnotes in the 
%abstract or reference list. Use letters for table footnotes.

%Unless there are six authors or more give all authors' names; do not use 
%``et al.''. Papers that have not been published, even if they have been 
%submitted for publication, should be cited as ``unpublished'' \cite{b4}. Papers 
%that have been accepted for publication should be cited as ``in press'' \cite{b5}. 
%Capitalize only the first word in a paper title, except for proper nouns and 
%element symbols.

%For papers published in translation journals, please give the English 
%citation first, followed by the original foreign-language citation \cite{b6}.

\end{document}